\documentclass[journal]{vgtc}
 
\usepackage[T1]{fontenc} \usepackage{tgtermes}

\usepackage{paralist}
\usepackage{listings}
\usepackage{enumitem}
\usepackage{amsmath}
\usepackage{amssymb}
\usepackage{cite}
\usepackage{graphicx}
\usepackage{url}
\usepackage[usenames,dvipsnames]{xcolor}
\usepackage[font=footnotesize,labelfont=bf,justification=centering]{caption}
\usepackage{subcaption}
\usepackage{float}
\usepackage{multirow}
\usepackage[normalem]{ulem}
\usepackage{array}
\usepackage{listings}
\usepackage{xspace}
\usepackage{pifont}
\usepackage{cleveref}
\usepackage{enumitem}
\usepackage{ wasysym }

\makeatletter
\def\url@leostyle{%
  \@ifundefined{selectfont}{\def\UrlFont{\sf}}{\def\UrlFont{\small\bf\ttfamily}}}
\makeatother
\setlist[itemize]{leftmargin=*}
\setlist[description]{leftmargin=*}

\def\pprw{8.5in}
\def\pprh{11in}

\usepackage[pdftex]{hyperref}
\hypersetup{
  pdftitle={SIGCHI Conference Proceedings Format},
  pdfauthor={LaTeX},
  pdfkeywords={SIGCHI, proceedings, archival format},
  bookmarksnumbered,
  pdfstartview={FitH},
  colorlinks,
  citecolor=black,
  filecolor=black,
  linkcolor=black,
  urlcolor=black,
  breaklinks=true,
}

\definecolor{amethyst}{rgb}{0.6, 0.4, 0.8}
\renewcommand{\paragraph}[1]{\vspace{7pt}\noindent\textbf{#1}}

\newcolumntype{C}[1]{>{\centering\let\newline\\\arraybackslash\hspace{0pt}}m{#1}}

\setitemize{noitemsep,topsep=0pt,parsep=0pt,partopsep=0pt}

\def\compactify{\itemsep=0pt \topsep=0pt \partopsep=0pt \parsep=0pt}
\let\latexusecounter=\usecounter

\newenvironment{CompactEnumerate}
  {\def\usecounter{\compactify\latexusecounter}
   \begin{enumerate}[leftmargin=*]}
  {\end{enumerate}\let\usecounter=\latexusecounter}

\title{Graphical Perception in Animated Data Visualizations}
\pdfoutput=1

\author{Eugene Wu, Lilong Jiang, Larry Xu, Arnab Nandi}
\authorfooter{
\item Eugene Wu is at Columbia University. E-mail: ewu@cs.columbia.edu
\item Lilong and Arnab are at Ohio State University.  E-mail: \{jianglil,arnab\}@cse.osu.edu
\item Larry is at U.C. Berkeley.  E-mail: larryxu@cs.berkeley.edu
}
\shortauthortitle{Wu \MakeLowercase{\textit{et al.}}: Perceptual Experiments of Animated Data Visualizations}

\abstract{ Interactive visual applications create animations that encode changes in the data.
For example, cross-filtering dynamically updates linked visualizations based on the user's continuous brushing actions.
The animated effects resulting from these interactions depends both
on how interaction~(e.g., brushing speed) controls properties of the animation such as frame rate,
as well as how the data that is being explored dictates the data encoded in the animation.
Past work has found that frame rate matters to general perception, however a critical question is
which of these animation and data properties affects the perceptual accuracy of judgement tasks, and to what extent.
Although graphical perception has been well studied for static data visualizations, it is ripe for exploration in the animated setting.
We designed two animated judgment tasks of a target bar in an animated bar chart and empirically evaluate the effects of 
$2$ animations properties---highlighting of the target bar and frame rate---as well as $3$ data properties that affect the target bar's value throughout the animation.
In short, we find that the rate and timing of animation changes is easier detected in larger values; that encodings such as color are easier to detect than shapes; and that timing is important---earlier changes were harder to perceive as compared to later changes in the animation.
Our results are an initial step to understanding perceptual accuracy for animated data visualizations, both for presentations and ultimately as part of interactive applications.

 }

\begin{document}
\maketitle


\section{Introduction}
\label{sec:intro}

Interactive visualizations are an integral part of visual analytics and a predominant method to explore larger, higher dimensional datasets.  From Shneiderman's classic introduction of dynamic queries in HomeFinder~\cite{williamson1992dynamic} to modern cross-filter~\cite{weaver2010cross} visualizations, these direct manipulation interfaces dynamically change the visualization in response to user interactions and enable users to rapidly generate and answer hypotheses about the data that cannot be answered by a static table or visualization~\cite{ehrenberg1975data}.  Both techniques can translate a single interaction (e.g., brushing motion) into rapid visualization updates and rely on the human's ability to accurately perceive information encoded in the resulting animation.

For instance, a cross-filtered visualization of flight delays may contain several views of the delay (by week, day, and hour). Dragging a brush in one of the views will update parameters of a filter (e.g., $hour\in [start, end]$) that translate into visual updates in each of the other views. In this case, each pixel of mouse movement defines the subset of data that is shown in an animation frame. Similarly, a user may move her mouse over the west coast of a map visualization to dynamically update a coordinated sales visualization and look for drastic or interesting changes across different counties. In both examples, each animation frame encodes a subset of the database controlled by filtering parameters controlled by the user interaction. This is in contrast to animated transitions~\cite{heer2007animated}, which synthesize animations between two separate visualization states for object tracking purposes. User decisions during an interactive session, such as what subset to analyze in detail, rely on making accurate perceptual judgements based on these animations.  However, despite the wide deployment and study of dynamic query interactions, the basic factors that affect the user's graphical perception of dynamically changing data visualizations is still under-explored. 

Existing graphical perception research has studied how numerous properties of the visualization, such as the visual encoding, the distractor bars, and the magnitude of the encoded data affect perceptual accuracy. The original Cleveland and McGill~\cite{cleveland1984graphical} studies measured the accuracy that users can perceive differences between pairs of marks under various visual encodings (e.g., length, position, angle). Follow-up work extended the design to crowdsourcing-based experimental platforms~\cite{heerperception}, studied separation and distractor effects~\cite{talbot2014}, and three-dimensional perspective~\cite{zacks1998}. Although these works have extended graphical perception in interesting directions, they have focused on pairwise comparison tasks, and static graphics; graphical perception in animated visualizations is ripe for exploration.

A natural question to ask is which factors of the animation, and to what extent do they, affect the accuracy of different perceptual tasks. For instance, does the frame rate always affect user accuracy, or only in certain cases? Are these effects independent of the data that is animated, or are there interactions between these parameters? In order to understand these questions, it is important to perform controlled, replicable studies and study the variation in user judgement accuracy in response to changes in these factors.

Prior studies have compared the effectiveness of animation with static representations of temporal data~\cite{robertson2008effectiveness} and found that animation is consistently less accurate. However, these studies do not control for the dataset nor the animation itself, and it is unclear what aspects of the animation contributed to the lower accuracy. Numerous perceptual experiments have studied the mechanisms that affect object tracking accuracy and have found that crowding~\cite{whitney2011visual} significantly degrades the ability to track objects, while other factors such as object speed and trajectory changes~\cite{tremoulet2000perception} appear to have minimal effects. Although these provide useful animation guidelines, and have been used to facilitate the design of animated transitions~\cite{heer2007animated}, is unclear to how to transfer and quantify the findings for animated data visualization.

This paper extends the graphical perception literature to animated data visualizations and studies the factors that affect perceptual accuracy. We designed and conducted a series of empirical experiments using animated bar charts that broadly surveys three important classes of factors.  The first class varies the complexity of the judgment tasks: a simple baseline task that asks users to estimate the maximum height of a target bar and a complex trend estimation task that asks users to reproduce how a target bar's value changed during the animation by estimating three properties of the trend. The second class of factors evaluates general animation properties such as frame rate and how the target bar is identified. The third class varies the data that is encoded in the visualization; this helps tease apart effects due to the animation as compared to the data being rendered. In addition to the contribution of our experimental design, we seek to address the following questions about perceptual accuracy:
1) how sensitive is perception to the individual factors, and under what conditions?
2) do the factors interact and in what ways?
3) when performing the complex task, how accurate are the individual sub-tasks?

We first present background and motivation, and an overview of our experimental design. We then validate our experimental setup by replicating the prior static visualization studies, report on our experiments, and conclude with implications of our findings as well as future work.

\section{Background and Motivation}
\label{sec:relwork}

The studies described in this paper are motivated by our desire to understand the perceptual implications
of interactive data visualizations.
As background, we first discuss two areas of closely related work---graphical perception studies stemming from the original
Cleveland \& McGill work~\cite{cleveland1984graphical} and studies of the efficacy and usage of animation 
in information visualization.
We finally describe interactive visualizations systems, the challenges of directly quantifying
the effects of interactions on user perception, and its relationship with animation.

Ultimately, a better understanding of perceptually accurate animated visualizations is 
important for interactive infoviz applications, where the analyst is dynamically sifting
through the dataset looking for and quantifying patterns of interest.  Understanding 
what will affect the user's conclusions is a crucial part of improving future interactive systems.

\subsection{Graphical Perception}
\label{s:rel-perception}

This paper builds upon theory and experimentation in the areas of psychophysics and graphical
perception~\cite{cleveland1984graphical,stevens1957psychophysical,falmagne1971generalized}
that study human perceptual accuracy and limitations in decoding visual encodings such as color and position.  
The most related are a series of graphical perception studies initially performed by Cleveland \& McGill.
They quantified the accuracy of pairwise comparisons in static data visualizations~\cite{cleveland1984graphical}
to gain insight into the effectiveness of different visual encodings.

In their design (Figure~\ref{f:static-examples}), users compare the heights of the bars labeled with circles.
They argue for this comparison task because \emph{the power of a graph is its ability to $\ldots$ see patterns and struture not readily revealed by other means of studying data},
and that \emph{conveying numbers with as many decimal places as possible} is best performed using tables~\cite{ehrenberg1975data}.
Heer et al.~\cite{heerperception} replicate these studies using a crowd-sourced design and a larger
variety of visualizations including pie charts and tree maps.
Talbot et al.~\cite{talbot2014} focus specifically on bar charts and study the effects of separation,
distractor bars, and participant bias towards multiples of $5$.
Zacks et al.~\cite{zacks2004using} study the effect of 3D perspective on bar charts.

These studies have followed Cleveland's original design and focused on static visualization and 
the same comparison tasks, however numerous challenges arise when translating the ideas to
animated visualizations.  First, the primary value of exploratory systems is that the exact analyses
are not known in advance, so that a static table is not appropriate.
Second, naively re-purposing the pairwise comparison task to an animation leads to ill-defined tasks with confounding factors.
For example, should the user report on how the comparison changes throughout the animation, or the comparison at a particular instant in the animation?  
The former approach confounds the effects of the animation due to the changes in the pair of bars, while the latter begs the question of which instant to choose.
Ultimately, new task designs are needed in order to quantify graphical perception for animations.

\subsection{Animation Studies}

Animation simulates continuous visual changes by the rapid sequence of static images.
Many original studies were based on principles from cartoon animation~\cite{chang1995animation} and 
motivated by applications to explain temporal data~\cite{rosling2009gapminder}, instructions~\cite{zongker2003creating,kehoe2001rethinking},
or state transitions~\cite{heer2007animated}.
Animation's general utility has received mixed reviews.
On one hand, it has been found to increase user engagement~\cite{wattenberg2006designing,tversky2002animation},
improve user orientation~\cite{tversky2002animation}, signal user attention
and help group objects through common fate~\cite{bartram1997can}.
On the other hand, unnecessary or ill-designed animation can also clutter the information
display, and reduce user comprehension.  In addition, animations may simply be difficult 
to perceive~\cite{tversky2002animation} and hard to accurately reproduce by the user~\cite{kaiser1992influence}

There is considerable evidence in the infovis community to use proper static visualization presentation techniques 
instead of animation for conveying changing data.
Tversky's survey~\cite{tversky2002animation} of animation research did not find a use of animation that outperformed a carefully designed static diagram.
When studying GapMinder-style animated bubble plots, Robertson et al.~\cite{robertson2008effectiveness} found that using animation to convey temporal
changes was less accurate and slower than static alternatives through the use of tracks or small multiples.
Although some users find animation to convey emotion or be more enjoyable, these are factors that are
useful for data presentation, and not necessarily for data exploration.

Animated transitions between two visualization states (e.g., swapping axes, or changing visualization encodings)
are a well studied use of animation that is intended to preserve object constancy~\cite{robertson1993information,palmer1999vision}.
Heer et al.~\cite{heer2007animated} develop a taxonomy of transition types that inform a set of 
animated data visualization transitions and find that a simple, staged approach reduces the error
in object tracking, and improves user understanding and engagement.
Chevalier et al.~\cite{chevalier2014not} present opposing results that staggered animated transitions have negligible utility.

Overall, prior animation research has provided insight into the psychological mechanisms for how we perceive dynamic visualizations,
and have provided a number of design guidelines for when to use animation, and what general properties, such as congruence and constancy that are
important to preserve.  However, there is opportunity to perform detailed studies akin to Cleveland's graphical perception studies, and quantify the sources
of perceptual accuracy, in animated visualizations.

\subsection{Interactions and Animation}

There has been considerable work on building highly interactive data exploration systems in the recent years.  
These systems, exemplified by Nanocubes~\cite{lins2013nanocubes} and imMens~\cite{liu2013immens},
provide interactive cross-filtering capabilities over million or billion-record datasets.
For example, imMens renders histogram summaries as bar and spatial map visualizations; as users
move their mouse over marks in the visualization, the charts are dynamically updated to reflect
counts of the selected subsets of the data.  
In effect, continuous user actions can be viewed as the cause of animated changes in a visualization.

In order to best utilize interactions, it is important to understand how users employ interaction and derive conclusions from the animated effects.
Interaction taxonomies~\cite{yi2007toward,heer2012interactive,keim2002information,lam2008framework} organize interactions based on 
e.g., high level goals, user intent, low level operations, and can help narrow the scope of interactions to evaluate.
User experiments have varied from qualitative studies of user interaction patterns~\cite{gotz2008empirical} to the effects of interaction delay~\cite{liu2014effects,gray2000milliseconds}
on user behavior and decisions.
However, detailed studies of the relationship between interaction, or even animation, and perceptual accuracy is still an emerging area;
it is unclear how to disentangle the numerous possible confounding factors in a direct interaction experiment.

Consider even a simple controlled design that uses a single interaction (e.g.,
a slider that dynamically filters a bar chart) and ask the user to perform a single judgement tasks. 
Even in this simplified case, are changes in perceptual accuracy due to the dataset (e.g., 
how the data changes, noise), the visualization (e.g., visual encoding, number of marks),
the animation (e.g., the frame rate), or the user behavior (e.g., where and how often the 
user chooses to scroll over a region)?

As it is clear that interaction will continue to be an integral part of information visualization, 
understanding how graphical perception is affected by dynamically changing visualizations is of importance.
Animation is arguably the most key visual feedback to the user's direct manipulation interactions and understanding
animation, on its own, affects perception will be a valuable precursor before turning our studies toward interaction.
\section{Study Design and Rationale}
\label{s:animated}

The goal of our study is to understand how a broad range of factors influence the accuracy of 
several judgement tasks in short animated visualizations.
Our basic task design renders a $2$-second bar chart animation and gives visual tests that require estimating values of a single
\emph{target} bar while ignoring other \emph{distractor} bars.
Many of the parameters, such as animation duration, are fixed throughout all experiments, while others, such as frame rate,
are tested.
We now discuss the rationale behind our task design and the choice of factors that we studied.

\begin{figure*}[t]
  \begin{subfigure}[b]{0.4\columnwidth}
    \includegraphics[width=0.90\columnwidth]{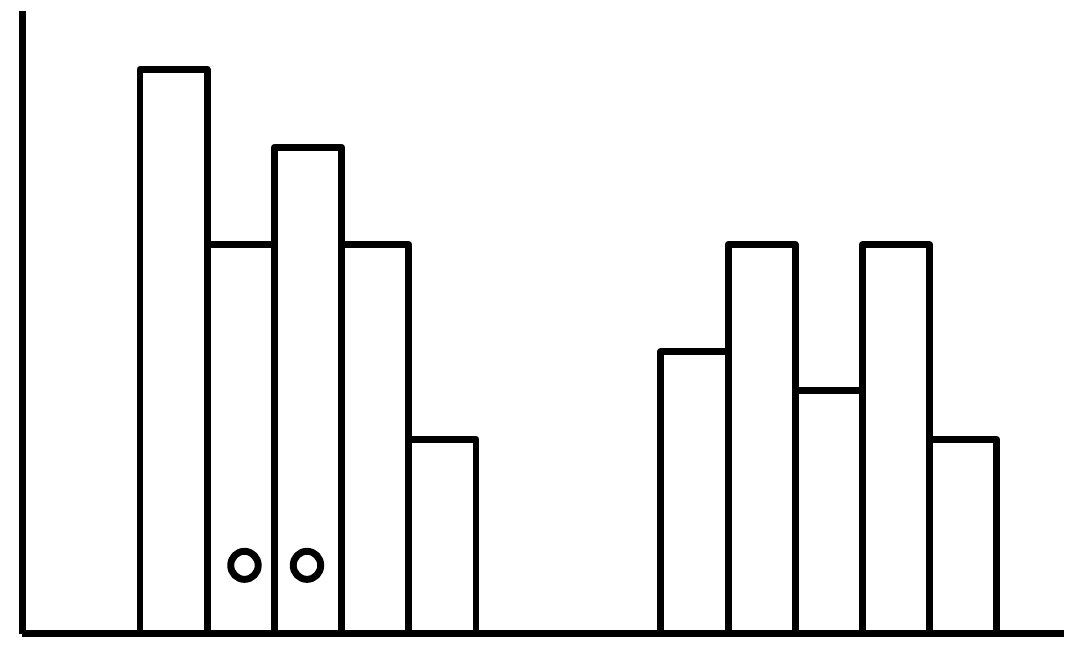}
    \caption{Adjacent Bars \\ (Type 1)}
    \label{f:static-adj}
  \end{subfigure}
    \begin{subfigure}[b]{0.4\columnwidth}
        \includegraphics[width=0.90\columnwidth]{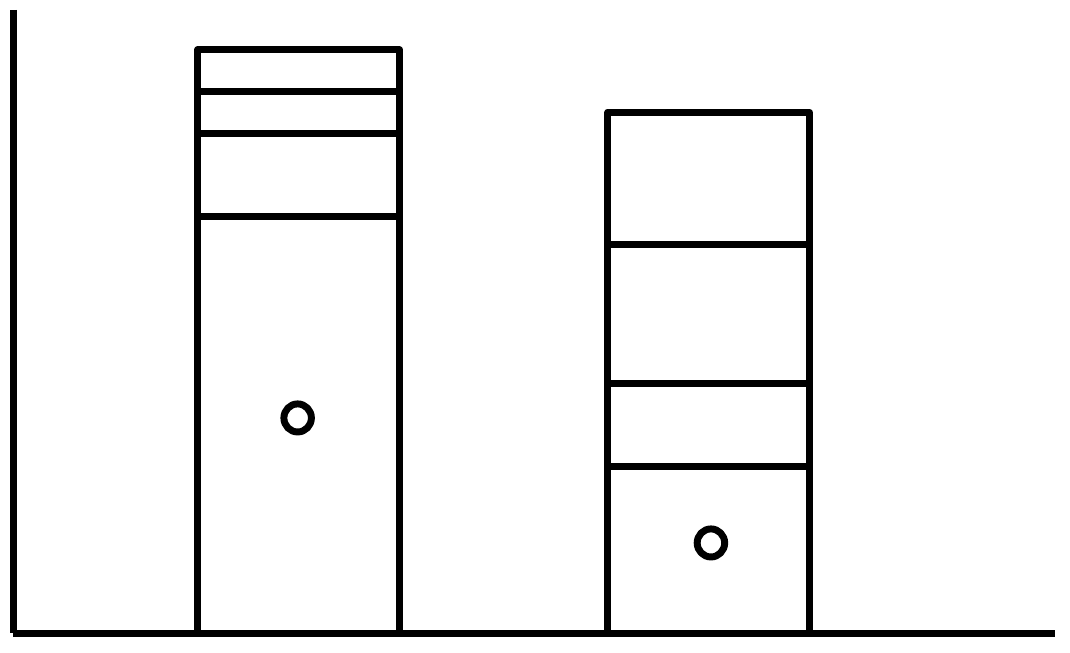}
        \caption{Aligned Stacked Bars \\(Type 2)}
    \label{f:static-alignstacked}
    \end{subfigure}
  \begin{subfigure}[b]{0.4\columnwidth}
	  \includegraphics[width=0.90\columnwidth]{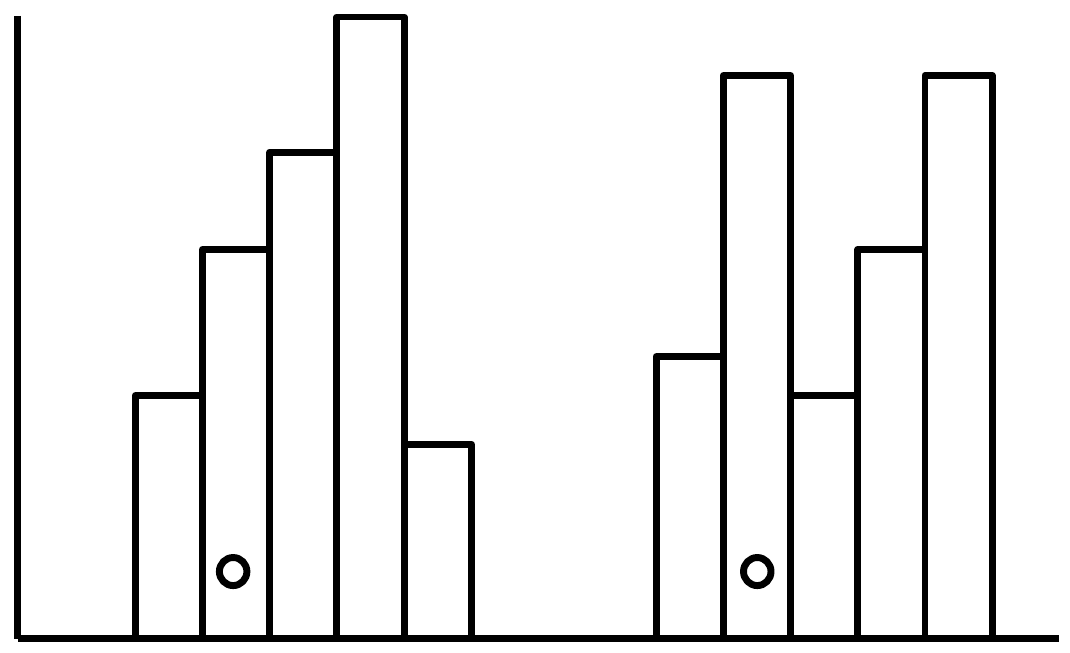}
	  \caption{Separated Bars \\(Type 3)}
    \label{f:static-sep}
  \end{subfigure}
  \begin{subfigure}[b]{0.4\columnwidth}
    \includegraphics[width=0.90\columnwidth]{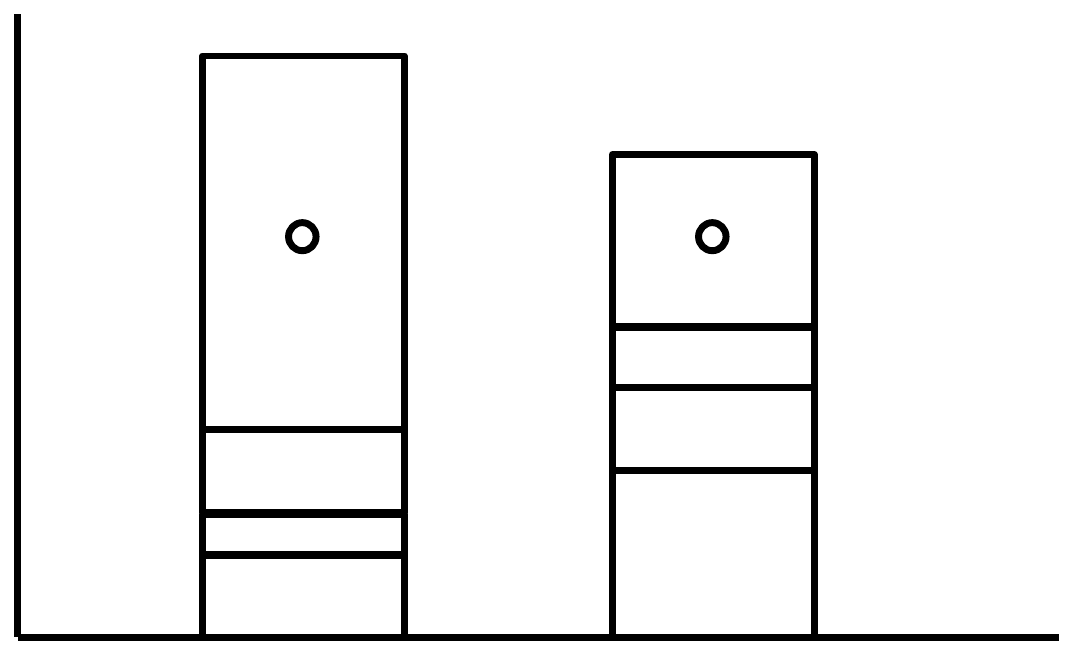}
    \caption{Unaligned Stacked \\(Type 4)}
    \label{f:static-unalignstacked}
  \end{subfigure}
  \begin{subfigure}[b]{0.4\columnwidth}
      \includegraphics[width=0.90\columnwidth]{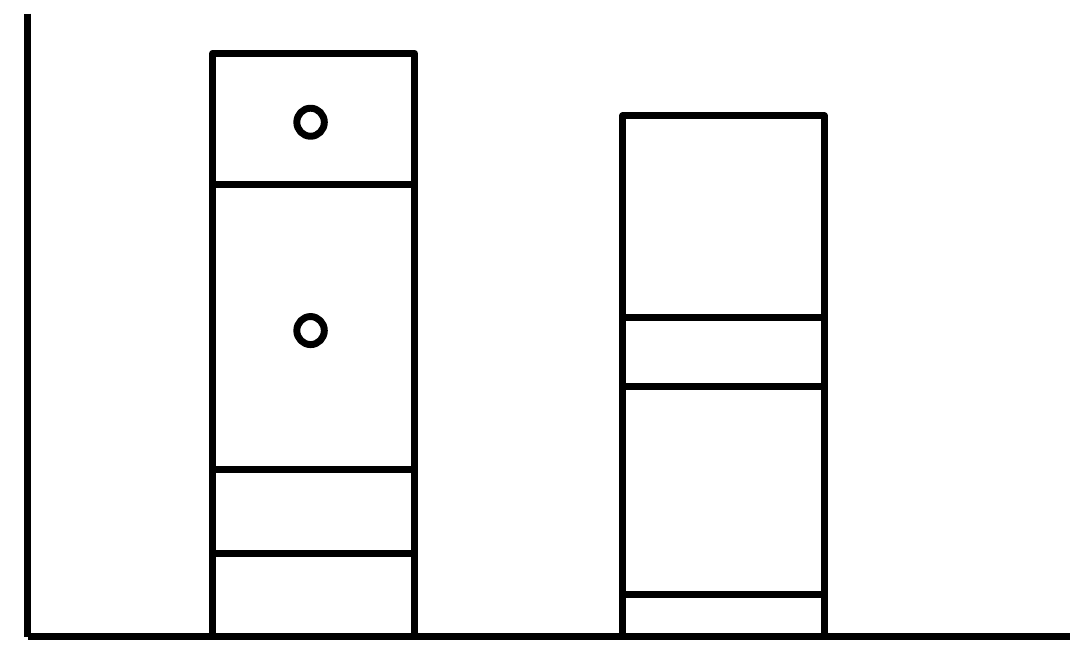}
      \caption{Divided Bars \\(Type 5)}
    \label{f:static-divided}
  \end{subfigure}
  \caption{Five types of bar charts used in static experiment.}
  \label{f:static-examples}
\end{figure*}

\subsection{Basic Task Design}
\label{s:taskdesign}

When permitted, we chose to remain consistent with  the design of prior graphical perception experiments~\cite{cleveland1984graphical,heerperception,talbot2014}.
We positionally encode the data as bar charts both because of its effectiveness as a visual encoding, and because it 
is well studied compared to alternative visualization types~\cite{heerperception,talbot2014,cleveland1984graphical}.
In prior static experiment, each chart showed a pair of target bars, along with four distractor bars for each target mark (see Figure~\ref{f:static-examples}).
To adopt this to our setting, we simplify the chart to contain the single target mark that the user compares across the animation frames,
and four distractors (Figure~\ref{f:animation}).

We assign random heights to the distractor bars in each animation frame~\cite{cleveland1984graphical,heerperception}
in order to avoid inadvertent correlated movements between the target and distractor bars, which may introduce a confounding factor~\cite{tversky2002animation}.
There is inconclusive evidence about how distractors affect judgement~\cite{talbot2014}, and we do not pursue this aspect in these experiments.
All charts are $380\times380$ pixels, without tick marks that may serve as reference points along the x and y axes.

We chose to show short animations based on a simplified model of user interaction
 $\mathcal{I} = [m_1, \ldots, m_N]$ as a piecewise linear sequence (possibly with gaps) 
of $N$ mouse movements $m_i$.  Each $m_i$ moves at a constant speed $s_i$ between the time intervals $[t^s_i, t^e_i]$.
For example, $m_i$ may model a single, short, linear scrubbing motion, and the user views the resulting animation $a_i$.
Our tasks focus on varying different properties of this animation unit $a_i$ and testing users's ability to accurately
perform judgment tasks.    

Our choice of the $2$-second duration is situated between Newell's~\cite{newell1994unified}
time scales for cognitive operations (e.g., clicking a link) and unit tasks (e.g., making a chess move).
This is because the tasks require users to not only identify patterns, which is a deliberate act on the order of 100ms, 
but also quantify exact amounts, which requires cognitive processing.

\subsection{Judgment Tasks}

As discussed in Section~\ref{s:rel-perception}, prior judgement tasks for static visualization studies
do not naturally translate to animated data visualizations due to the additional temporal setting.
For instance, if the user is asked to perform pair-wise comparisons between a pair of bars, what does the user report?
How the comparison changes throughout the animation?  Or the largest difference between the pair?

As Cleveland~\cite{cleveland1984graphical} noted, comparison tasks are fundamental to the user of visualizations.
Thus, we focus on the comparison of a single bar's value \emph{across animation frames}, 
and ask users to perform judgement tasks that measure different characteristics of the 
change of a single target bar's value.    
For this, we chose two tasks from Yang's taxonomy~\cite{yang} of task-oriented visual insights that are used in trend analysis and 
range from simple to complex.

The baseline \texttt{extrema} task asks users to estimate the value of the target bar at its turning point~\cite{Bianchi1999}.
Turning points are important in trend analysis because they signify an abrupt and significant change in the trend that can be a candidate for detailed investigation.
For the patterns that we study (Section~\ref{s:setup-params}), this is equivalent to asking users to estimate the maximum value $v_{max}$ of the target bar in the animation~\footnote{
We didn't study judgments of minimum value to keep the number of parameters manageable.  }.
This task is of comparable complexity as prior static studies.

The complex \texttt{trend} task is commonly used~\cite{robertson2008effectiveness} to characterize the overall shape of the trend
(e.g., estimate the target bar's values that are defined by the curves shown in Figure~\ref{f:trendcontrols}).
It is more complex because it requires the user to estimate multiple aspects of the trend such as when it changes, the bar height at the turning point, and how quickly it changes.
The bar height estimate is the same as the one in the baseline extrema task, and provides an opportunity to compare the accuracy between the two judgement types.

\subsection{Parameters Tested}
\label{s:setup-params}

Animations can be described along two separate dimensions: the \emph{data} that is encoded in the animation frames (data parameters)
and how the animation is \emph{styled and rendered} (animation parameters).  
The data dimension defines what is shown, while the animation dimension defines how the animation is styled and played back.
We study the effects of both types and describe them below; they are summarized in Table~\ref{t:params-wide}.

\begin{table}[t]
\hspace*{-.1in}
\centering
\small
  \begin{tabular}{l|l|l|l}
    {\bf Parameter} & {\bf Type} &  {\bf Description} & {\bf Defaults} \\ \hline
    $mark\in \{\circ, {\color{red}{\blacksquare}}{\color{blue}{\blacksquare}}\}$  & Anim  & Target bar's marking.  & color \\
    $fps\in [2, 30]$   & Anim  & Frames per second.  & $10$ \\   \hline
    $v_{max}\in [22, 85]$ & Data & Maximum value. & {\tiny $22, 32, 47, 69, 85$}\\
    $pos\in [0,1]$     & Data & Maximum Position. &  $0, 0.5, 1$ \\  
    $k\in [1, 100]$       & Data & Rate of change. & $1$ \\ \end{tabular}
  \vspace*{-.1in}
  \caption{Parameters tested in experiments}
  \label{t:params-wide}
\end{table}

\subsubsection{Data Parameters:} 
\begin{figure}[htb]
  \centering
  \begin{subfigure}[b]{0.32\columnwidth}
    \includegraphics[width=\columnwidth]{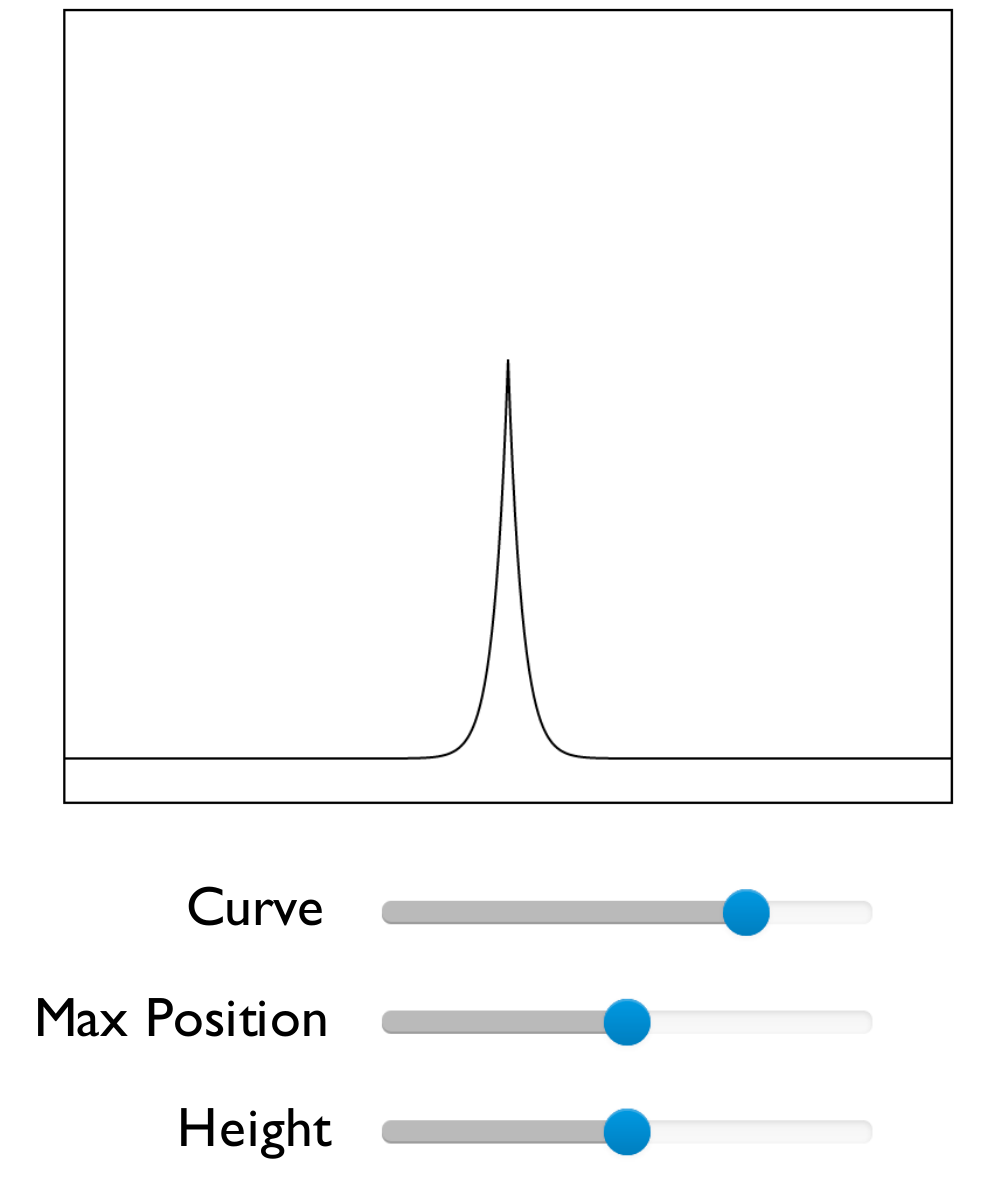}
    \vspace*{-.1in}
    \caption{High rate of change ($k$)}
    \label{f:tc4}
  \end{subfigure}
  \begin{subfigure}[b]{0.32\columnwidth}
    \includegraphics[width=\columnwidth]{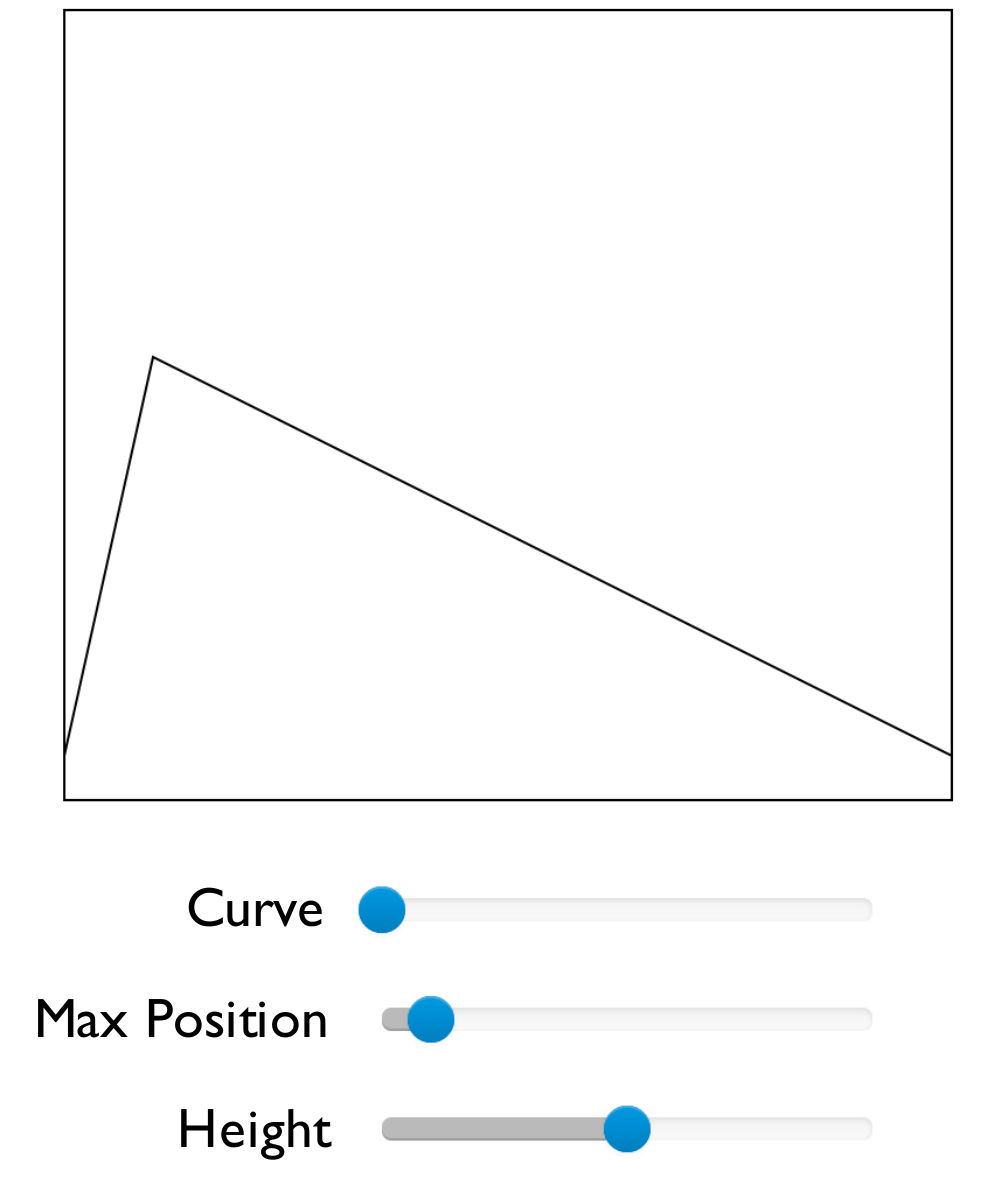}
    \vspace*{-.1in}
    \caption{Low Max Position ($pos$)}
    \label{f:tc1}
  \end{subfigure}
  \begin{subfigure}[b]{0.32\columnwidth}
    \includegraphics[width=\columnwidth]{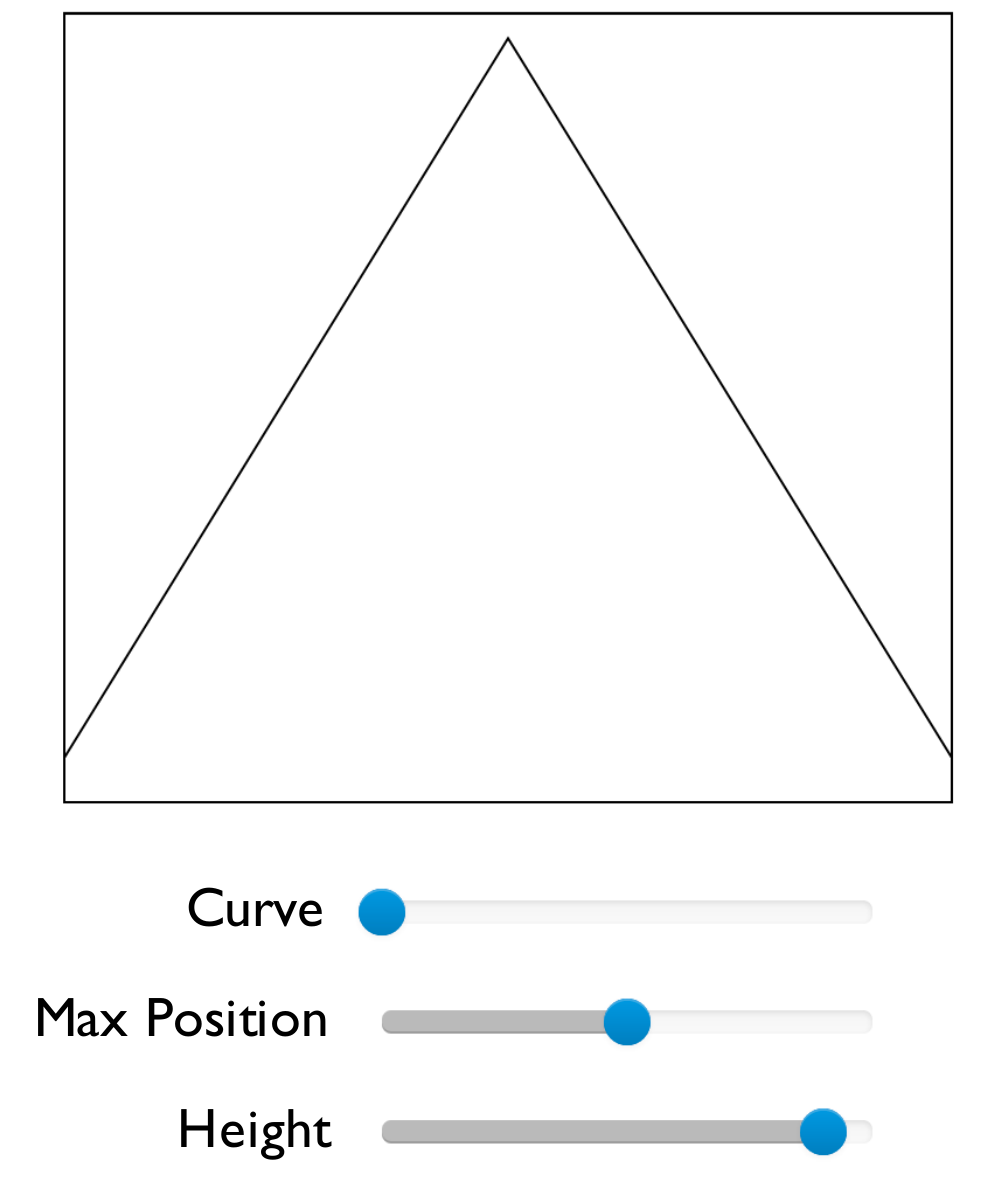}
    \vspace*{-.1in}
    \caption{High max value ($v_{max}$)}
    \label{f:tc3}
  \end{subfigure}
	\vspace*{-.1in}
  \caption{Examples of three trend functions along with user input controls for the trend judgement task.}
  \label{f:trendcontrols}
	\vspace*{-.1in}
\end{figure}

The data parameters define the data that is rendered during the animation.
One approach was an ecologically valid design that uses real datasets (e.g., world bank data) to generate the animations~\cite{robertson2008effectiveness}.
The main limitation is the inability to isolate individual factors such as the magnitude of the values or how the values change over time.
We instead used a simple data generation model---the distractor bars are generated randomly (Section~\ref{s:taskdesign}) and the target bar's height is defined by a trend function $\mathcal{T}$.
The trend function must be designed to support a wide variety of patterns yet only require a small number of parameters.
We found that a simple template---variations of a pyramid-shaped pattern containing a single turning point (Figure~\ref{f:trendcontrols})---revealed interesting insights about the two judgement tasks.
Furthermore, complex trend patterns are composed of sequences of turning points, so our results may serve as a baseline future experiments.
The key limitation is the difficulty of isolating what parts of the datasets ultimately that affect user judgements.

The trend function $\mathcal{T}(x)$ maps the position of a frame $x \in [0, 1]$ in the animation to the target's value.
This function can generate variations of a pyramid-shaped pattern by tuning its three parameters: $v_{max}$, $pos$ and $k$.
The target bar increases from an initial value $v_i$ to a maximum value $v_{max}$, and then decreases to $v_e$.
The frame containing the bar that reaches $v_{max}$ is called the maximum position $pos \in [0, 1]$
and varies from the first to the last frame (Figure~\ref{f:tc1},\ref{f:tc3}).
Additionally, trends change at different rates---inflation increases at a constant rate, however
the number of edges in a social network increase non-linearly.  The exponential term $k$ models this aspect by controlling the rate of change.  
When this value is low, the function follows a linear trend, however large values such as
$100$ approximate an impulse function that has low values at all points except for the frame corresponding to 
the position parameter $pos$,  where the target value reaches the maximum value $v_{max}$ (Figure~\ref{f:tc4}).
{
\begin{align*}
\mathcal{T}(x) = \begin{cases}
      (v_{max}-v_i) \frac{x}{pos}^{k} + v_{i}       & \mbox{if } x \le pos \\
      (v_{max}-v_e)  \frac{1 - x}{1-pos}^{k} + v_{e} & \mbox{otherwise}
                 \end{cases}
\end{align*}
}
Finally, trends typically exhibit smaller perturbations in addition to the overall trend shape.
To replicate this, we add a small amount of gaussian noise $\mathcal{N}(0, 5)$ to the target bar.
In a follow-up study, we specifically analyze the effects of various noise models on user perception
in animated visualizations and find that user accuracy is surprisingly robust to the amount of noise.
However we use the simple noise model in this paper.

We are careful to ensure the reported errors are due to perceptual inaccuracies rather than an artifact of the experimental design. 
When varying data and animation parameters, we ensure that the target bar is always within $[5, 95]$, and that
the maximum value of the target bar is exactly $v_{max}$ in a single frame of the animation as defined by the $pos$ parameter.
We also ensure that the inclusion of random noise does not cause the target bar 
to exceed or equal $v_{max}$ in any other frame.

\begin{figure}[ht]
  \begin{subfigure}[b]{0.32\columnwidth}
    \includegraphics[width=0.95\columnwidth]{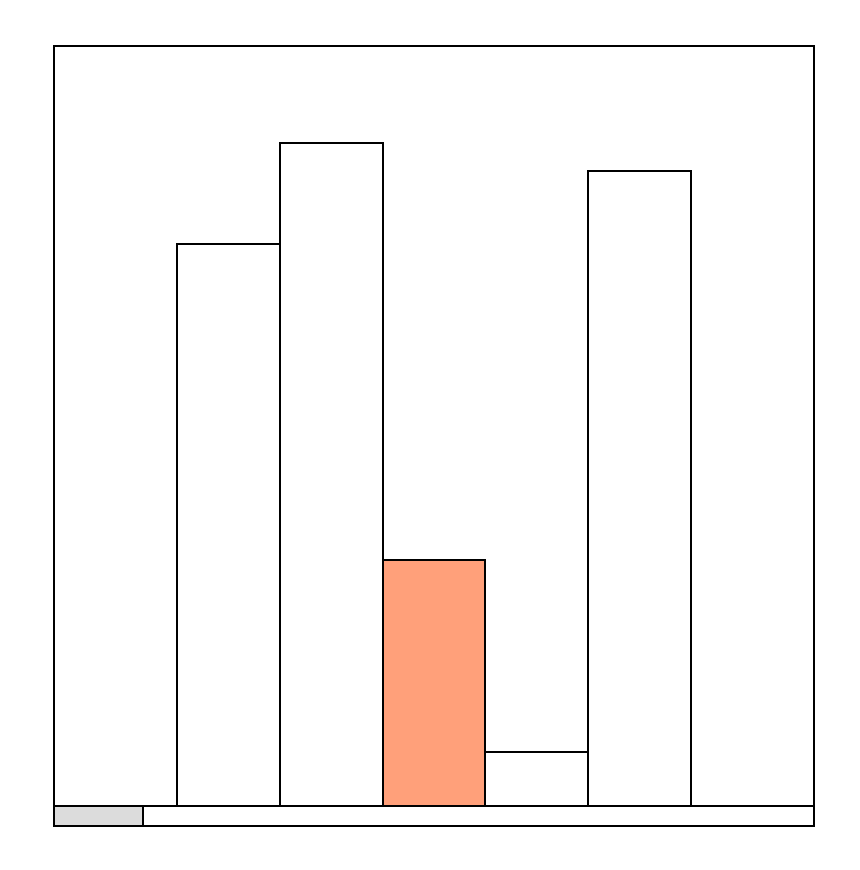}
    \caption{10\% done, target marked with color}
    \label{f:anim1}
  \end{subfigure}
  \begin{subfigure}[b]{0.32\columnwidth}
      \includegraphics[width=0.95\columnwidth]{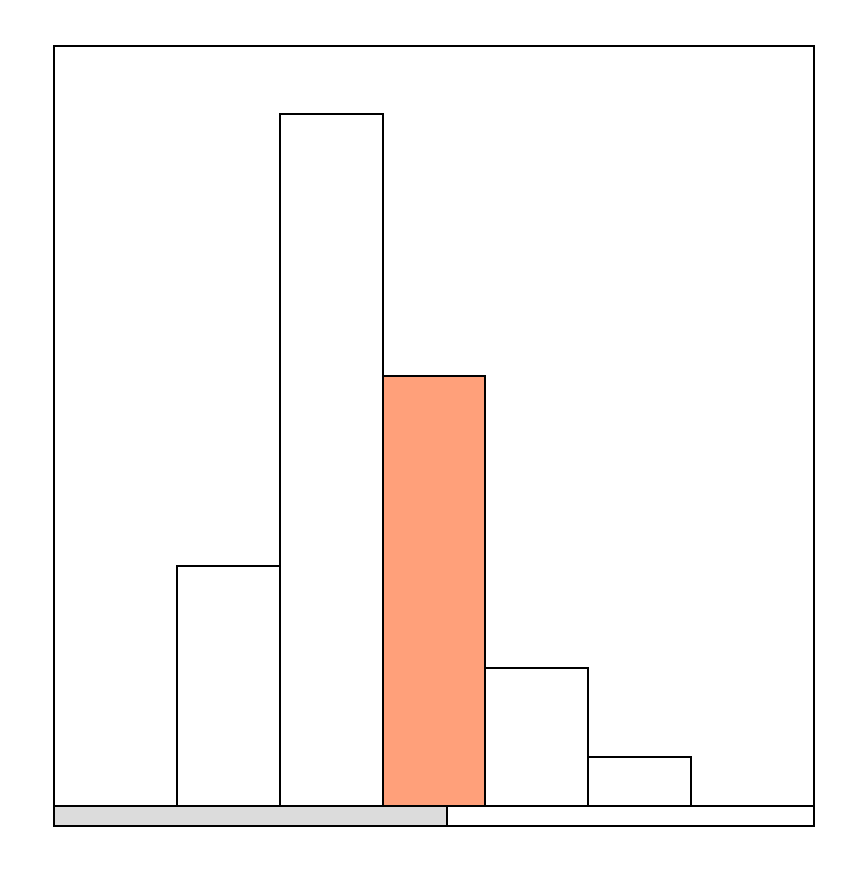}
      \caption{50\% done, target marked with color}
    \label{f:anim2}
  \end{subfigure}
  \begin{subfigure}[b]{0.32\columnwidth}
      \includegraphics[width=0.95\columnwidth]{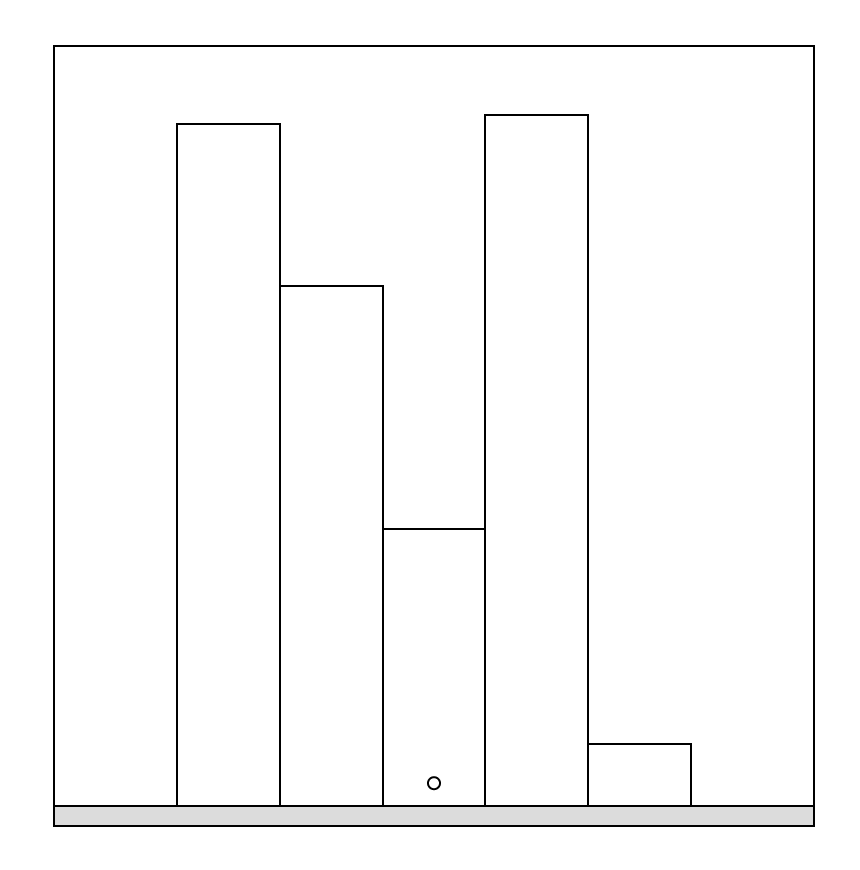}
      \caption{100\% done, target marked with circle}
    \label{f:anim3}
  \end{subfigure}
  \vspace*{-.1in}
  \caption{\small Examples of frames under different progress and marking conditions.
    The grey bar at the bottom shows the animation progress.  
    The target bar is marked with color in \ref{f:anim1},~\ref{f:anim2}, and circle in \ref{f:anim3}.}
  \label{f:animation}
\end{figure}

\subsubsection{Animation Parameters:} 
The prior graphical perception studies used a circle shape to mark the target bars.
The preattentive literature suggests that color identification exhibits ``pop-out''
that improves tracking performance~\cite{healey2012attention}.
For this reason, we study the effects of marking the target bar ($mark$) using a circle and a colored fill.

In addition, we simulate different speeds of the mouse movement $s_i$ (Section~\ref{s:taskdesign})
by varying the frames-per-second ({\it fps}) of the animation.     
Because we fix the animation to $2$-seconds, so changes in the frame rate
alter the number of frames sampled from $\mathcal{T}$ but not the duration.
An alternative design that fixes the number of samples and varies the animation
duration is a direction for future studies.

\subsection{Experimental Overview: }
We employ a within-subjects format and a custom multi-task interface to manage the multiple judgment tasks in each HIT.
We track user progress so users with intermittent connectivity can return to the last incomplete task.
Each experiment is a separate HIT deployed on Amazon Mechanical Turk and we pay subjects using bonuses. 
Similar to previous studies, subjects must pass several qualification and training tasks before participating in an experiment.  
We ensure worker comprehension of the task interface by using drastically simplified versions of the tasks as qualification and rejecting 
subjects that fail the qualification.
The qualification task is designed specifically for the experiment and described in the corresponding section.  
In addition, an initial qualification task renders an animation at 30 frames per second (fps) and terminates 
the assignment if the true frame rate is below 27 fps;  we also track the true fps during an assignment.  
These measures ensure that the judgments are not biased by the browser's capability (e.g., a slow browser may render the animation slower than intended.)
 
The training tasks are the same format as the actual tasks, however we show the correct
answer upon submission and do not reject subjects at this point to avoid bias.  
Workers are presented judgments in random order, and
we record auxiliary information such as the browser, display size, and response time.
Workers were paid $5\cent$/judgement, including qualification, 
or $\$18-30$/hr ($5.5$s/extrema, $10$s/trend judgement).

Prior to analysis, we filter out incomplete, spammers and outlier responses.
We remove spammers whose responses are all nearly identical, based on the difference 
between maximum and minimum estimates (less than 15 for extrema experiments, less than 50 for trend experiments) and standard derivation of estimates (less than 10).  
We remove workers whose responses have less than a $0.8$ correlation with the true values~\cite{talbot2014}.
Finally, we remove individual responses $3$ standard deviations outside of the response mean.  
Our findings are robust to the specific threshold values.

Log transformation is regularly used in models of human performance~\cite{limpert2001} to help
address skewed residuals that appear in linear models~\cite{kaybeyond}.
We similarly compute the user's absolute error $\epsilon_{v_{max}} = |\textrm{judged val - true val}|$ 
and compute its log transform~\cite{cleveland1984graphical} $\log_2(\epsilon_{v_{max}} + \frac{1}{8})$,
and report the $25\%$ trimmed means along with $95\%$ bootstrapped confidence intervals~\cite{talbot2014}.
We varied the trim percentage to $15\%$ and found our results were robust to this setting.

Due to the unexplored nature of this area of research, we were forced to make a trade off between practical budget limitations and coverage of experiments -- we chose to prioritize experimental coverage.
For this reason, many of our experiments reflect sample sizes from a modest number of assignments due to many disqualified workers.
Despite this, confidence intervals are kept relatively tight in many experiments, and we are still able to draw insights from the results.

The subsequent sections describe the three overarching experiments that we conducted.  
Section~\ref{s:static} validates our crowd-sourced experimental setup by reproducing prior 
static graphical perception studies.
We then extend the design to animated graphics for the extrema task in Section~\ref{s:extrema},
and the trend characterization task in Section~\ref{s:trend}.

\section{Experiment 1: Validating Crowdsourcing Setup}
\label{s:static}

We first demonstrate the consistency of our experimental setup with existing
work by re-running prior crowdsourced protocols~\cite{talbot2014,heerperception}.

\subsection{Materials and Procedure}

We ask subjects to estimate the height ratio between two target bars in 5 types of bar charts 
borrowed from Cleveland \& McGill~\cite{cleveland1984graphical} (Figure~\ref{f:static-examples}).
Each chart shows the two bars in addition to additional {\it distractor} bars.
We study 9 true percentages between the two bars: $18, 22, 26, 32, 39, 47, 55, 69, 85\%$.
We ensure that at the same ratio, the compared bars have the same height across all five types of charts. 
The heights of the randomized distractor bars are constrained such that, for a given true percentage, they are 
the same between the adjacent and separate chart types (e.g., Figures~\ref{f:static-adj} and~\ref{f:static-sep}),
and between the aligned and unaligned stacked charts (e.g., Figures~\ref{f:static-alignstacked} and~\ref{f:static-unalignstacked}).

The prior experiments distinguished the target bar by placing a circle or dot in the bar.
However, gestalt theory~\cite{wertheimer1944gestalt} 
suggests that a colored target bar (e.g., Figure~\ref{f:anim1}) would be easier to distinguish from the distractors due to preattentive ``pop-out'' effects---this may have an effect when the bar chart is animated.
Although this experiment uses a static image, we additionally test this factor in order to serve as a comparison point for the animation experiments.

The qualification task uses a multiple choice selector where one of the choices is obviously correct.
Similar to Talbot~\cite{talbot2014}, we remove incomplete workers and those whose answer 
correlation with the true percentage is less than $0.8$. 

\begin{figure}[H]
  \centering
  \vspace*{-.1in}
  \small
  \includegraphics[width=.75\columnwidth]{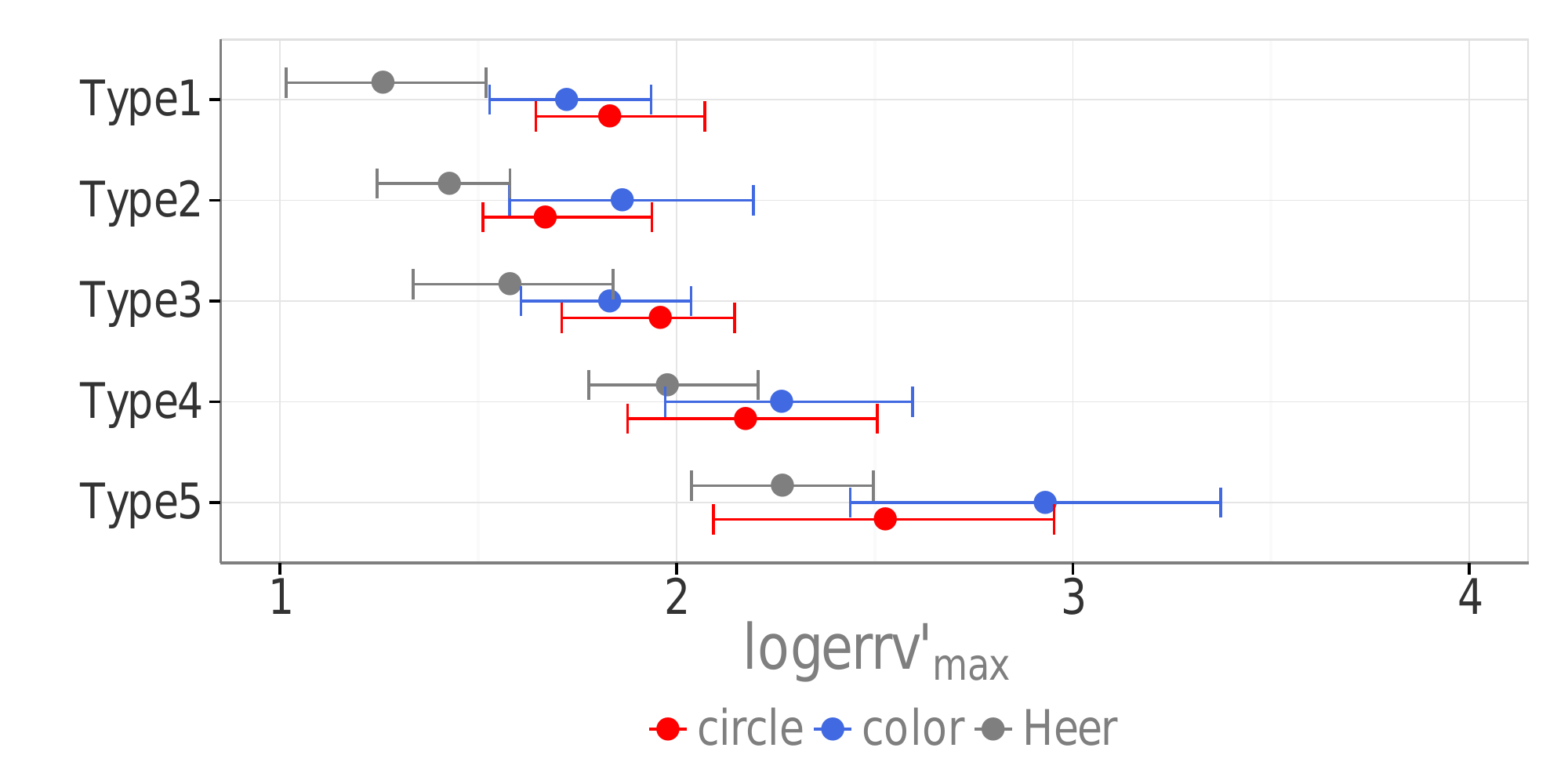}
  \vspace*{-.1in}
  \caption{95\% conf. intervals for different bar chart types}
  \label{f:static_conf}
\end{figure}

\subsection{Results}

We analyzed $47$ and $48$ out of $54$ completed assignments that passed the correlation threshold 
for circled and colored markings, respectively.
The resulting rankings (Figure~\ref{f:static_conf}) are comparable to both the Cleveland \& McGill 
and Heer \& Bostock experiments, although the distinctions between each chart type is less pronounced than prior studies.  
For example, adjacent and aligned stacked bar charts showed similar magnitudes of judgment errors.  
It is expected that the circle and color mark types exhibited no statistically significant differences
in a setting where timing and motion were not a factor.

\section{Experiment 2: Reading Extrema Values}
\label{s:extrema}

Our first animated experiments study user perception of extrema values in animated bar charts. 
Our key goals are to understand:
\begin{CompactEnumerate}
\item How is accuracy affected by the target's marking?
\item Under what conditions does frame-rate affect accuracy?
\item How is accuracy affected by the data that is rendered?
\end{CompactEnumerate}

\subsection{Procedure}

\begin{figure}[ht]
  \centering
  \includegraphics[width=0.5\columnwidth]{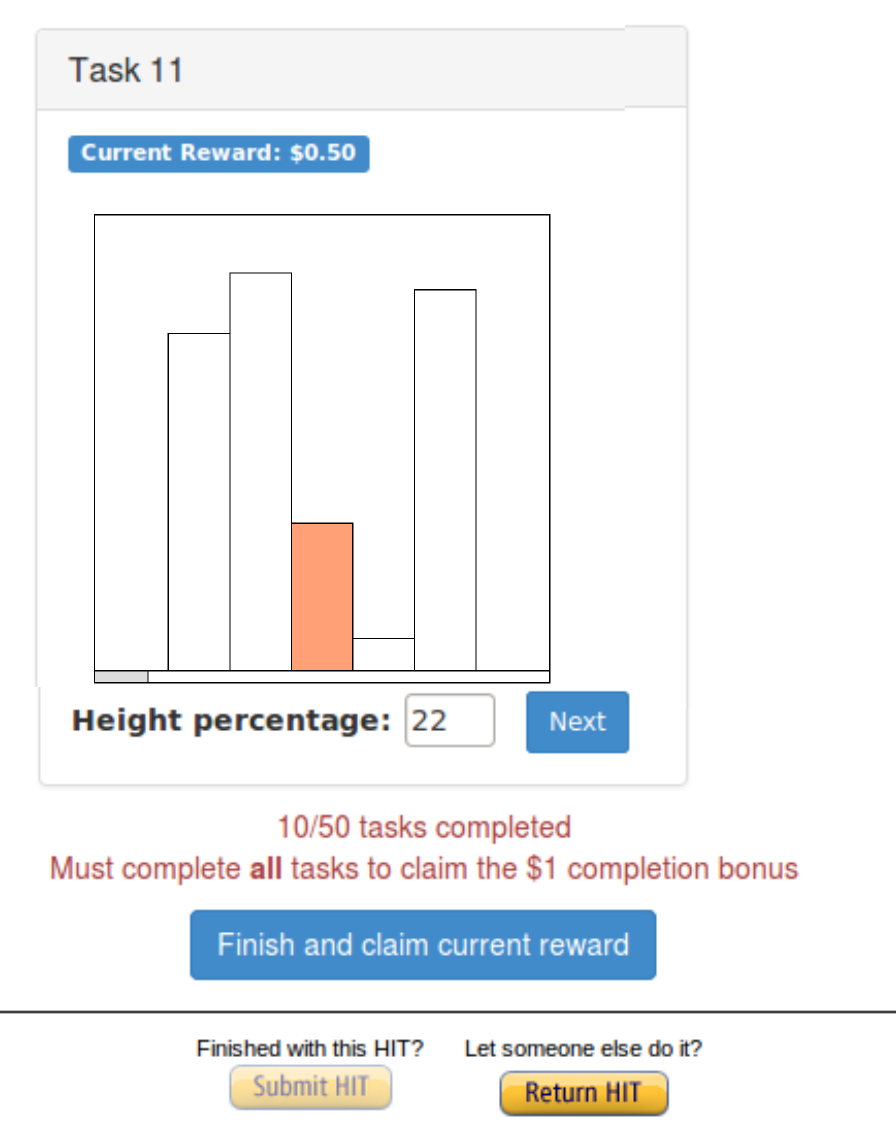}
  \caption{\small Example interface for extrema task.  User is completing the $11^{th}$ judgement, and will be given a $\$1$ bonus if she completes all judgements. }
  \label{f:taskinterface}
\end{figure}

Users were asked to make ``quick visual judgements''.
When the user clicks the start button, the button is replaced with a three second countdown to ensure that the user is ready.
We then render the animation once, along with a gray progress bar at the bottom, and hide the bar chart when the animation ends so that it is only shown once.
The worker is then asked to estimate ``the maximum height reached by the marked bar as a percentage of the total chart height.''---this estimate is denoted $v_{max}'$.
The in-progress animation interface is shown in Figure~\ref{f:taskinterface}.

The qualification tasks in these experiments use a multiple-choice version of an animation task
where $v_{max} > 70$.  The choices include three low values and one correct high value.
We then showed three sample training tasks that showed the answers upon submission.

We ran an initial experiment using a coarse parameter sweep, and used the results to design the three subsequent experiments.
We studied worker responses and found that their errors did not systematically improve as they completed more tasks.
This suggests that they are not improving through prior experience, and we thus 
allow repeated workers across experiments (though only allow them to complete a given experiment once).
Table~\ref{t:animated-params} summarizes the parameter values we used in all animated experiments in this paper,
along with the number of factors in the design.  

\begin{table*}[t]
  \small
  \centering
  \begin{tabular}{c | c | c | c | c | c  }
  			& {\it fps} & {\it mark} 	& {\it pos} 	& {\it k} &   \\ 
  {\bf Experiment} 					& {\bf Frame Rate} & {\bf Mark} 	& {\bf Maximum Position} 	& {\bf Rate of Change} &   {\bf \# Factors} \\ \hline
  \texttt{Initial-study}   			 & 2, 10, 30     & color, circle  & 0, .5, 1 & 1     & 45   \\ 
  \texttt{Extrema-fps$|$k}       & 2, 10, 20, 30 & color  & 0, .5, 1  & 1, 5, 10 & 180 \\
  \texttt{Extrema-k}             & 30	           & color, circle  & 0, .5, 1  & 1, 2, 5, 10, 25, 50, 100 & 105  \\
  \texttt{Extrema-pos}           & 10            & color  & 0, .1, .15, .25, .4, .5, .6, .75, .8, 1 & 1, 5 & 100 \\
  \texttt{Trend}            		 & 10, 30        & color  & 0, .1, .25, .5, 1 & 1, 5, 10, 25 & sampled 100 of 200  \\
  \end{tabular}
  \caption{Parameter configurations used in animated experiments.}
  \label{t:animated-params}
\end{table*}

\subsection{Results}\label{s:extrema-results}

We fit the data using a log-linear model that estimates the log-error of the user
estimate $y_{f,u}$ (e.g., $v_{max}'$)---dependent on the estimated parameter value $f$
(e.g., $pos$) and the user $u$---as a linear combination of the input parameters $p_{i,u}$ 
(e.g., $p_{pos,u}$, $p_{v_{max},u}$).
The user specific error $\mathcal{U}_u$ is assumed to follow a normal distribution with standard
deviation $\sigma_f$ dependent on $f$.
The coefficient $\beta_{f,i}$ is the model's sensitivity to the $i^{th}$ parameter, dependent on $f$:
\vspace*{-.1in}
\begin{align*}\small
log(y_{f,u}) &= \beta_{f,0} + \sum_{i=1}^n \beta_{f,i}*p_{i,u} + \mathcal{U}_u\\
\mathcal{U}_{u} &\sim \mathcal{N}(0, \sigma_f)
\end{align*}
In the text, we use the term \emph{error} or \emph{absolute error} to refer to $\epsilon_{v_{max}}$, 
and \emph{log error} to refer to its log transformed value.

\begin{figure}[h]
\centering
\vspace*{-.1in}
\includegraphics[width=0.9\columnwidth]{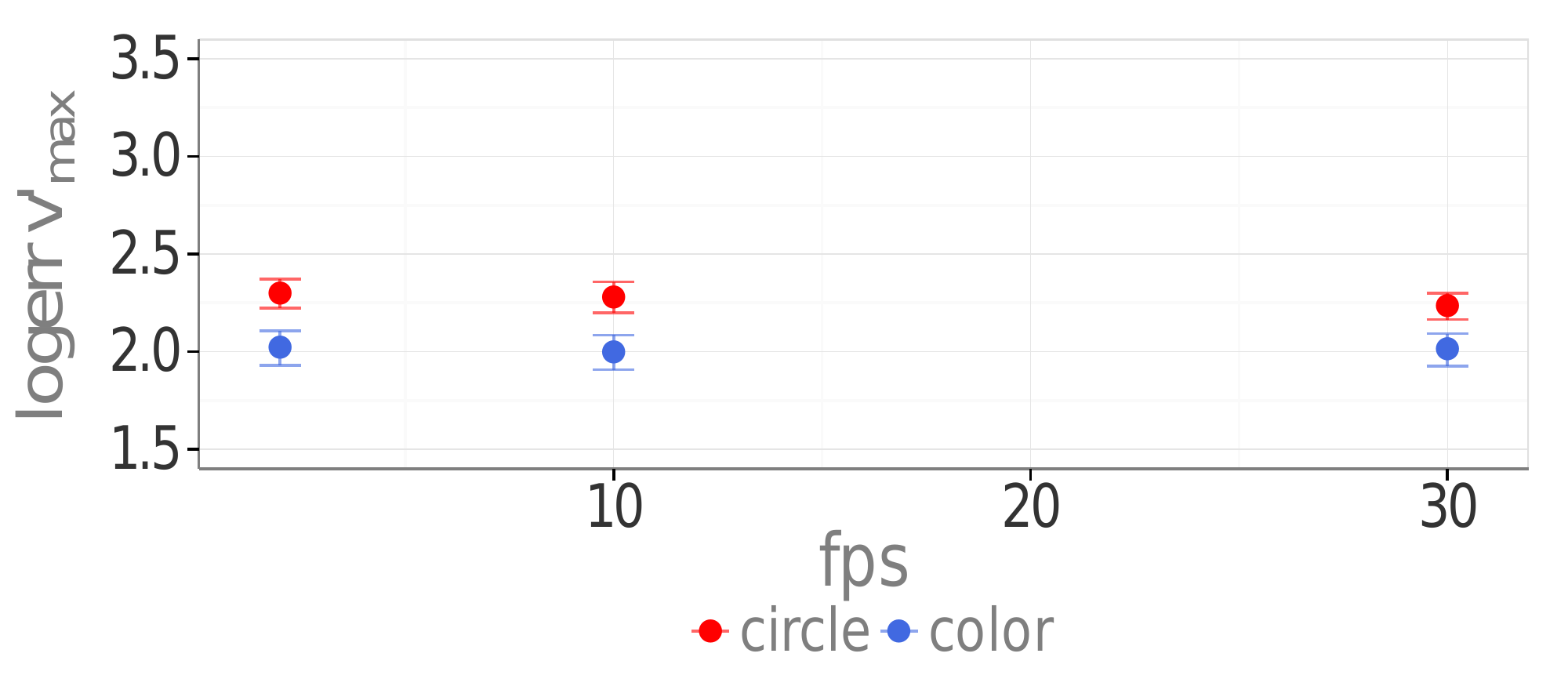}
\vspace*{-.1in}
\caption{Log error vs frame rate for {\color{red}{circle}} and {\color{blue}{color}} markings.}
\label{f:init_color_vs_circle}
\vspace*{-.1in}
\end{figure}

\smallskip
\noindent\textbf{Initial Study (\texttt{Initial-study}): } 
Our initial study performed a broad parameter sweep across the frame rate, maximum position, the max value,
and the target marking (see \texttt{Initial-study} in Table~\ref{t:animated-params}).
We ran separate HITs for each marking type.  There were $35$ and $20$ completed assignments, 
$29$ and $17$ assignments after removing spammers and outliers, for \emph{circle} and \emph{circle}, respectively.
An unpooled t-test on log error found a significant effect of \emph{mark} ($t(1584.4)=5.336, p = 1e^{-7}$) 
with an effect size of $0.404$---this is equivalent to $\sim32\%$ difference in absolute error. 
These results are consistent with gestalt theory~\cite{wertheimer1944gestalt} which predicts that color 
is easier to perceive than shape due to ``pop-out'' than shape (Figure~\ref{f:init_color_vs_circle}).

We then fit the log-linear model to gauge how the accuracy depends on the input parameters.  For the circle marking,
we found a negative relationship to $pos$ ($\chi^2(1,N=1285)=7.9, p=.005, \beta_{pos}=-0.286$).
The color marking exhibits a slightly stronger negative relationship with $pos$
($\chi^2(1,N=759)=6.11, p=.01, \beta_{pos}=-0.356$).  This means that the log error decreases by $\beta_{pos}$
when the maximum position is at the end of the animation instead of the beginning.

The log-linear fit suggests that peaking later in the animation is slightly easier to detect.  
We were surprised to find that frame rate did not have a significant impact on the perceived accuracy, 
which is contrary to common sense expectations.
We hypothesize that this is because the linear trend was easily predictable and 
makes the judgment task simple regardless of frame rate.  
For this reason, we used the rate of change $k$ parameter and studied non-linear rates of change.

\smallskip
\noindent\textbf{Frame Rate and Rate of Change (\texttt{Extrema-fps$|$k}): }
In this experiment, we focus on the color marking and increase the granularity of the frame rate $fps$ and rate of change $k$.
This results in a $180$ factorial design---the large number of judgements 
resulted in $16$ complete assignments, and $15$ after filtering.

\begin{figure}[h]
\centering
\includegraphics[width=0.9\columnwidth]{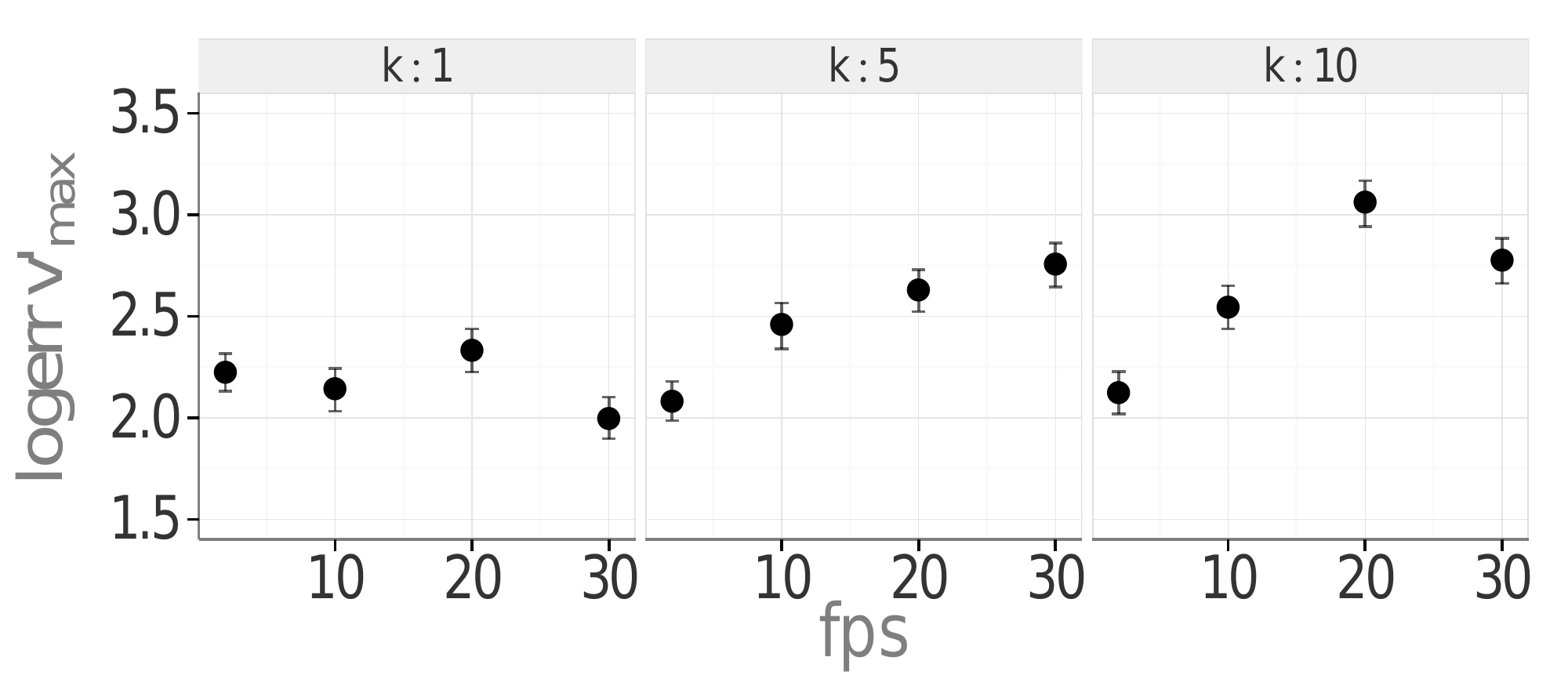}
\vspace*{-.1in}
\caption{\small Log error vs frame rate, conditioned on rate of change $k$}
\label{f:logerr_fps_k}
\vspace*{-.1in}
\end{figure}

Figure~\ref{f:logerr_fps_k} plots log error against frame rate for each rate of change $k$,
and shows that higher frame rates result in higher estimation error, particularly when $k\ge5$.
On average, there is a strong dependence on $k$ ($\chi^2(1,N=2684)=34, p=5e^{-9}, \beta_{k}=0.05$), 
meaning that increasing $k$ by $1$ increases the absolute error by $3.5\%$.
There is a significant dependence on $fps$ ($\chi^2(1,N=2684)=38.1, p=6e^{-10}, \beta_{fps}=0.018$),
so that increasing the frame rate by $10$ increases absolute error by $\sim13\%$.  
We then studied the dependencies conditioned on $k$ and found that there is no dependence on frame rate
when $k=1$. However, when $k=5,10$, the absolute error increases by nearly $22\%$ for every increase of the frame rate by $10$ ($\beta_{fps} = 0.029$). 
This is likely because larger $k$ values more closely approximate an impulse function, and a larger $fps$ will show the peak for a shorter duration.

\begin{figure}[h]
\centering
\includegraphics[width=0.9\columnwidth]{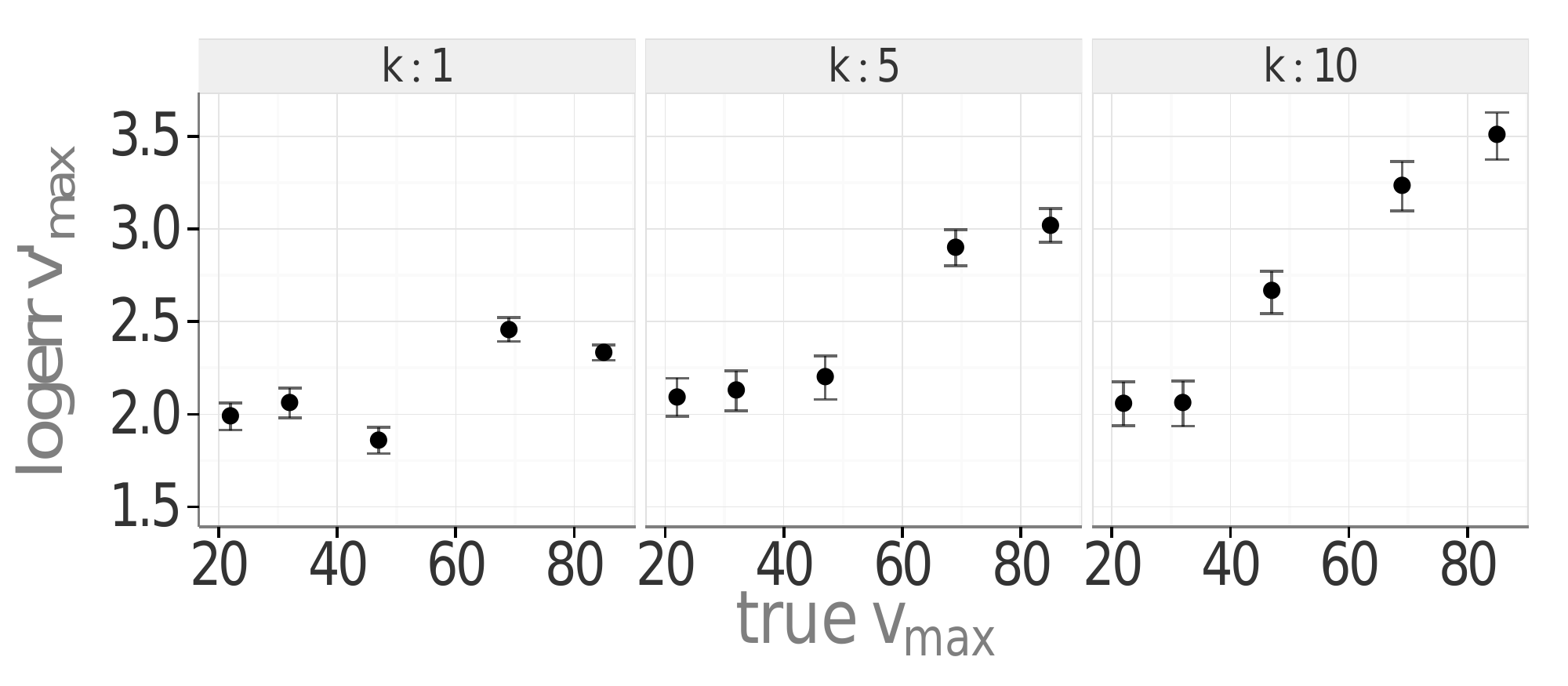}
\vspace*{-.1in}
\caption{\small Log error vs true $v_{max}$, conditioned on rate of change $k$}
\label{f:logerr_truth_k}
\vspace*{-.1in}
\end{figure}

Figure~\ref{f:logerr_truth_k} plots log error against the true $v_{max}$ for each $k$.
Prior studies~\cite{cleveland1984graphical} found that perceptual accuracy depends on the magnitude of the judgement,
and we find a similar dependency.  The log error when the true $v_{max} < 40$ appears stable, however rapidly increases for higher values.
In addition, the rate that the log error increases grows with the target bar's rate of change $k$.
When $k=10$, the log error when the true height is $85$ is nearly $2.8\times$ larger than when it is $22$

\smallskip
\noindent\textbf{Rate of Change (\texttt{Extrema-k}): }
\label{ss:extremapower}
Given the strong dependence on rate of change $k$, we further increased its granularity
while fixing the frame rate at $fps=30$.
In addition, we re-evaluated the effect of the mark types in the context of $k$. 
Each subject completed $105$ judgments and we received 
$30$ and $28$ complete assignments for circle and color markings, respectively.

Initially, we found that nearly all of the participants were removed by the correlation threshold as part of spammer and outlier removal.
We found that the accuracy degrades by nearly $3\times$ as the rate of change increases from $5$ to $50$ and contributed to the low correlation.
We concluded that the task is intrinsically difficult when $k\ge25$.
Instead, we only considered user judgements when $k\le20$ when applying the correlation-based filter.
We ultimately analyzed $20$ and $27$ assignments for the circle and color markings.

\begin{figure}[h]
\centering
\vspace*{-.1in}
\includegraphics[width=0.9\columnwidth]{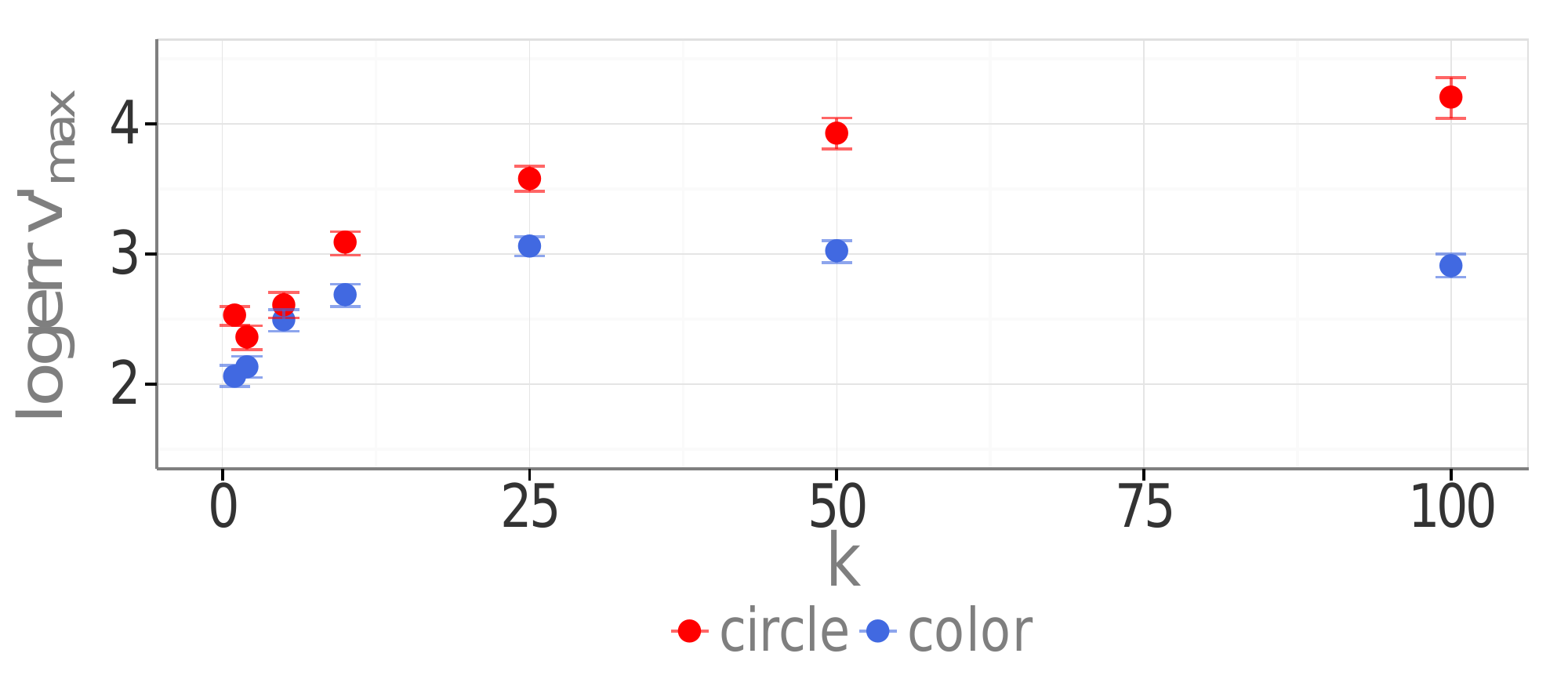}
\vspace*{-.1in}
\caption{Log error vs rate of change.}
\label{f:logerr_exponent}
\vspace*{-.1in}
\end{figure}

Figure~\ref{f:logerr_exponent} shows that accuracy quickly degrades as the rate of change increases, 
and converges to a maximum error rate as $k>25$ for color marking, when the trend appears to approximate an
impulse function. 
We found a statistically significant difference between color and circle ($t(4429.3)=9.855, p<2e^{-16}$) that is consistent with the initial study.
Applying ANOVA, we found a significant effect of $k$ for both circle
($\chi^2(1,N=2085)=239, p=6e^{-54}, \beta_{k}=0.016$) and color markings 
($\chi^2(1,N=2813)=95.1, p=2e{-22}, \beta_{k}=0.009$).  
These results motivate our focus on the range of $k\in[0, 25]$ in the subsequent experiments.

\smallskip
\noindent\textbf{Maximum Position (\texttt{Extrema-pos}): }
\label{ss:extremapos}
This experiment studies sensitivity to the {\it pos} parameter at a finer granularity, 
conditioned on smaller rates of change.
We used $10$ levels for $pos$, reduced $k$ to $\{1, 5\}$, fixed $fps = 10$, and used color marking.
Each subject completed $100$ judgments, and we analyzed $45$ out of $48$ complete assignments.

\begin{figure}[h]
\centering
\vspace*{-.1in}
\includegraphics[width=0.9\columnwidth]{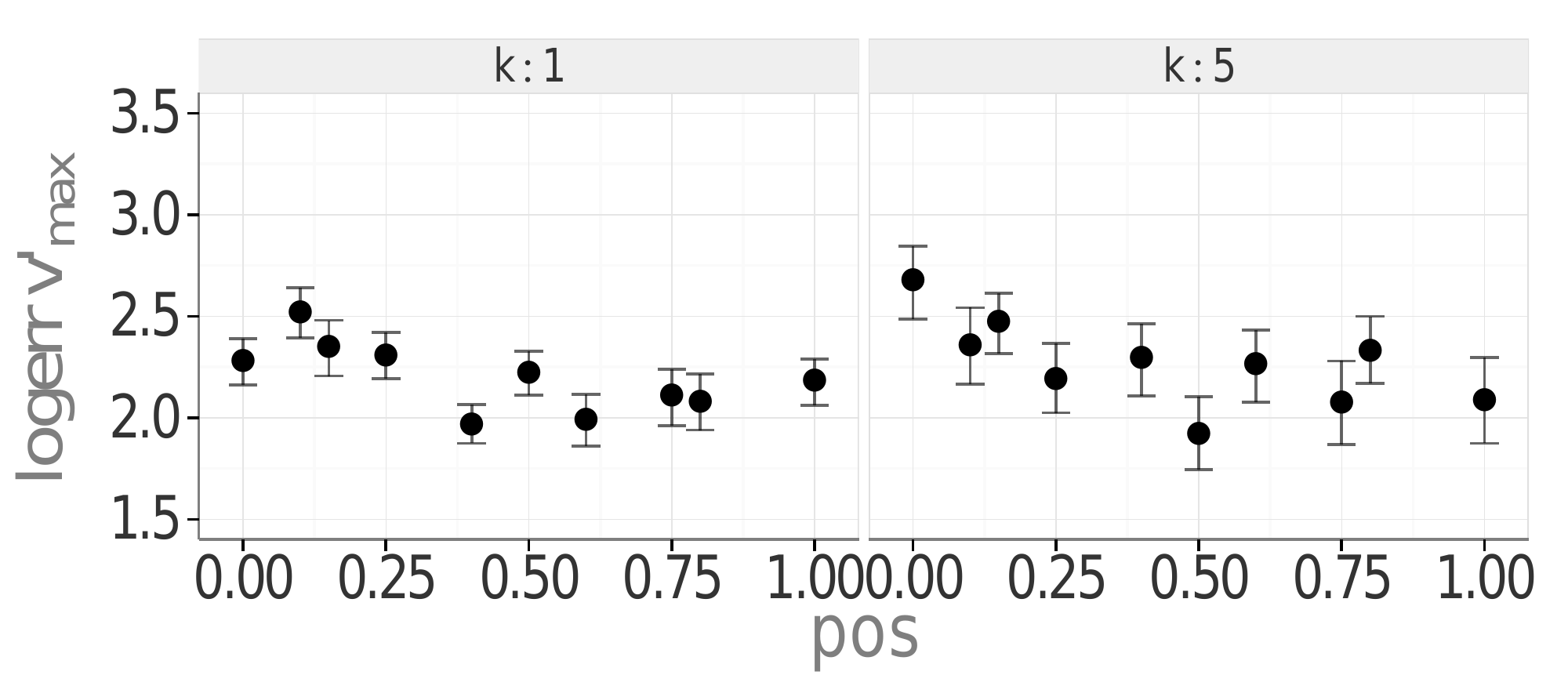}
\vspace*{-.1in}
\caption{\small Log error vs max position, conditioned on rate of change.}
\label{f:logerr_maxpos}
\vspace*{-.1in}
\end{figure}

Figure~\ref{f:logerr_maxpos} compares log error against $pos$ for each $k$ value.
On average, these is a statistically significant negative relationship with $pos$ 
($\chi^2(1, N=2215)=14.9,p=1e^{-4}, \beta_{pos}={-0.4}$), verifying that peaks later in the 
animation are easier to perceive.
Conditioned on $k$, the negative relationship is much stronger with higher $k$:  
the $\beta_{pos}$'s magnitude increases from $-0.374$ to $-0.437$ as $k$ increased from $1$ to $5$.
Despite the countdown to prime the participant before the animation runs, it is possible that this effect
is an artifact of the short animation and may disappear if a longer animation is used.

\smallskip
\noindent\textbf{Discussion: } 
Our results reinforce the hypothesis that using color to distinguish the target bar is more effective than using shape.
We also find that perceptual accuracy depends on both data and animation parameters. 
When the rate of change is linear, accuracy is robust to varying frame rates---if this holds true for other forms of animated judgement, 
it suggests possibilities in reducing the frame rate, and consequently, the computational costs when the data changes linearly.
On the other hand, larger rates of change, even $k=5$, both reduces judgement accuracy, and increases the sensitivity to the other parameters.
Given the dependence on rate of change, interactions that vary the speed of the animation in order to ensure that the target bar changes at a linear rate
may improve readability of the extrema when the temporal dimension is not a factor in the task.
Further studies are needed to validate these claims and ensure additional biases are not introduced from such manipulation.

\begin{figure}[h]
\centering
\vspace*{-.1in}
\includegraphics[width=\columnwidth]{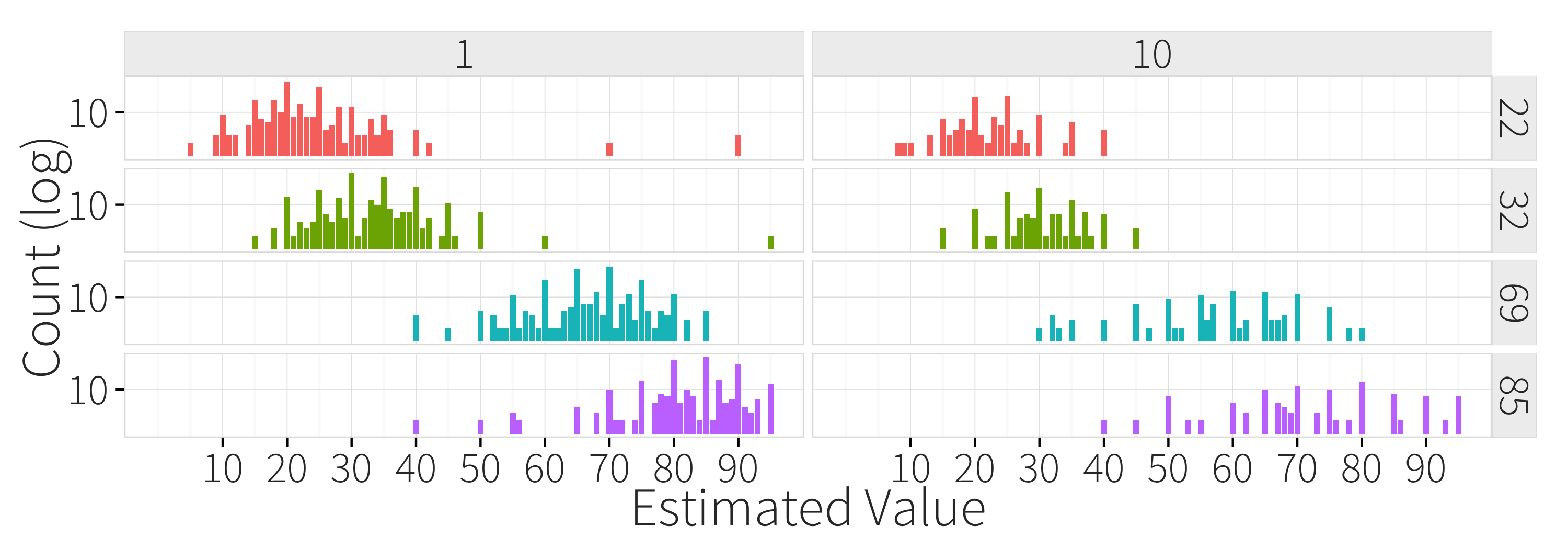}
\vspace*{-.1in}
\caption{\small Histogram (log scale) of judgments for varying rate of change (columns) and true max heights (rows).}
\label{f:estmax_counts}
\vspace*{-.1in}
\end{figure}

We found that although log error depends on the true height $v_{max}$, Figure~\ref{f:logerr_truth_k} showed that
log error appeared constant when $v_{max}<40$.  This assymetric distribution is a departure from prior height comparison studies~\cite{talbot2014}
where absolute error peaked when the true percentage was $40$. 
We plotted a histogram (log scale) of all responses from experiments with $pos\in\{0, .5, 1\}$, $30$ fps, and color marking.  
Figure~\ref{f:estmax_counts} is facetted by the rate of change along the columns and four true $v_{max}$ values by row.
We first note the user bias towards multiples of $5$ seen by Talbot et al.~\cite{talbot2014}.
We also find that the distribution of estimates for smaller values is relatively clustered around the true value,
while there is a long tail of very small estimates for larger $v_{max}$ values (e.g., 85).   
This is most evident when $k=10$.
We verified that the long tail is not related to worker quality, and hypothesize that users are simply more 
likely to miss the frames containing the maximum value and report a severe underestimate.

A consequence of the experimental design is that the frame containing $v_{max}$ is always shown, regardless of the frame rate.
In practice, say when the user scrubs a slider at different rates, we cannot guarantee the frame will be shown, and expect lower perceived accuracy.

\section{Experiment 3: Distribution Characterization}
\label{s:trend}

\begin{figure}[tb]
  \centering
  \includegraphics[width=.7\columnwidth]{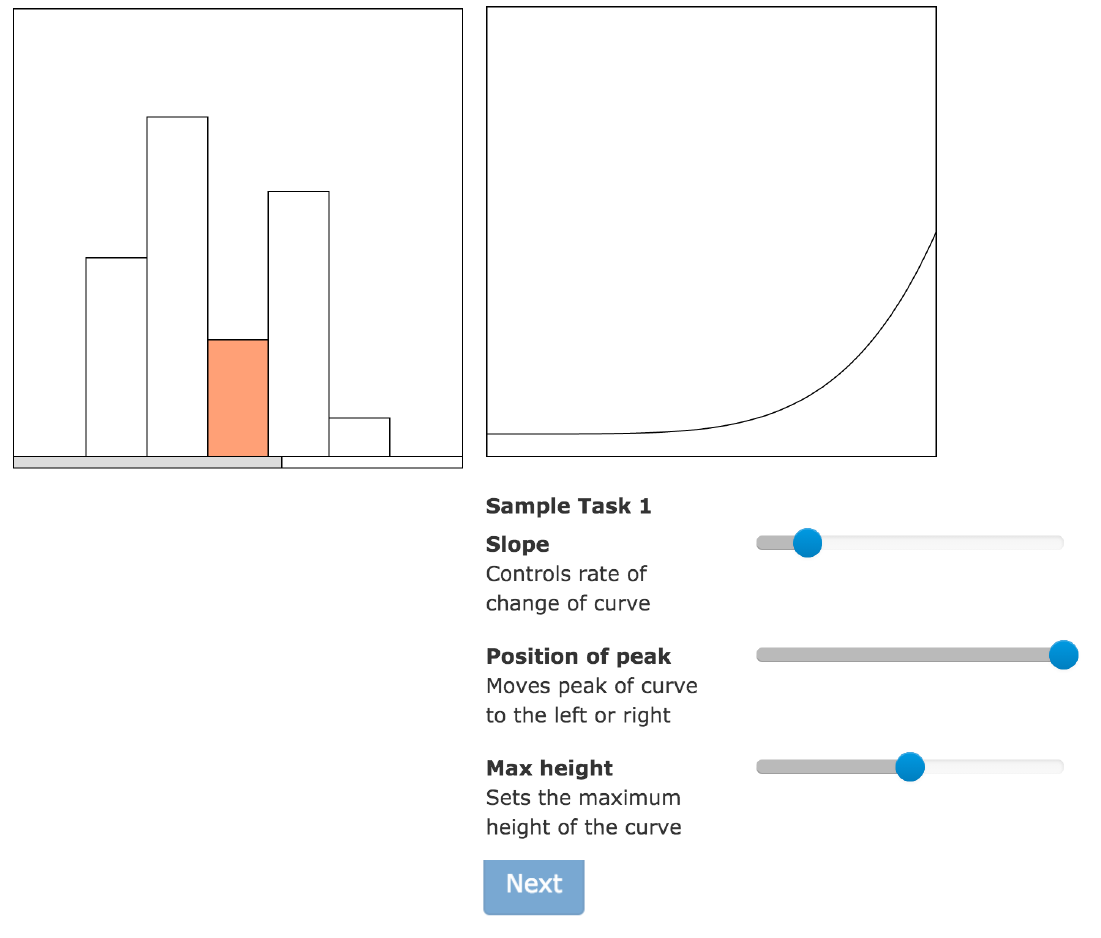}
  \vspace*{-.1in}
  \caption{Example interface for trend judgement task.  The right side shows user controls 
  and visualized trend function. }
  \label{f:trendinterface}
\end{figure}

\begin{figure*}[tbh!]
  \centering
  \begin{subfigure}[b]{0.24\textwidth}
    \includegraphics[width=1\columnwidth]{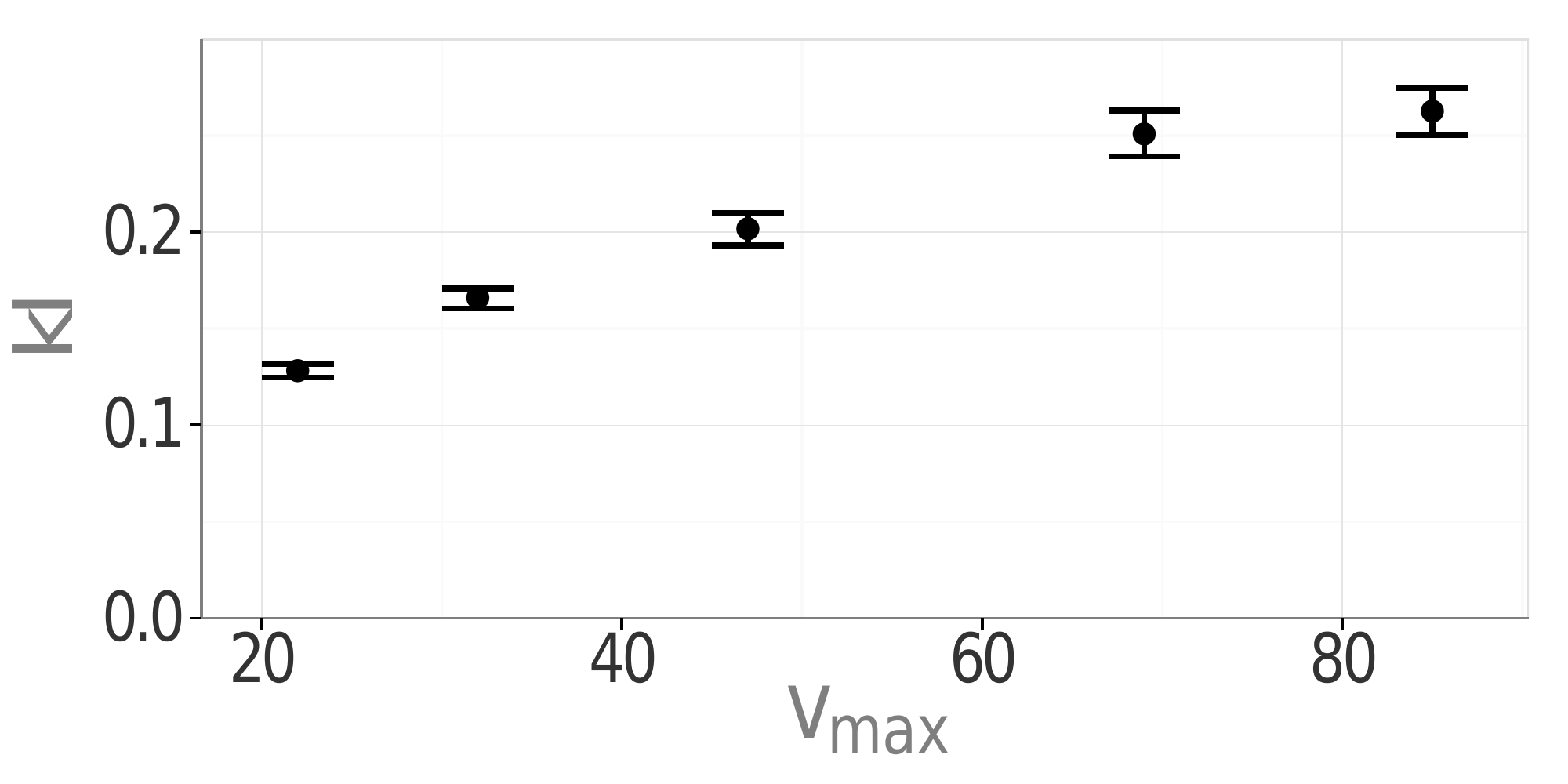}
    \vspace*{-.15in}
    \caption{$kl$ vs $v_{max}$}
  \end{subfigure}
  \begin{subfigure}[b]{0.24\textwidth}
      \includegraphics[width=1\columnwidth]{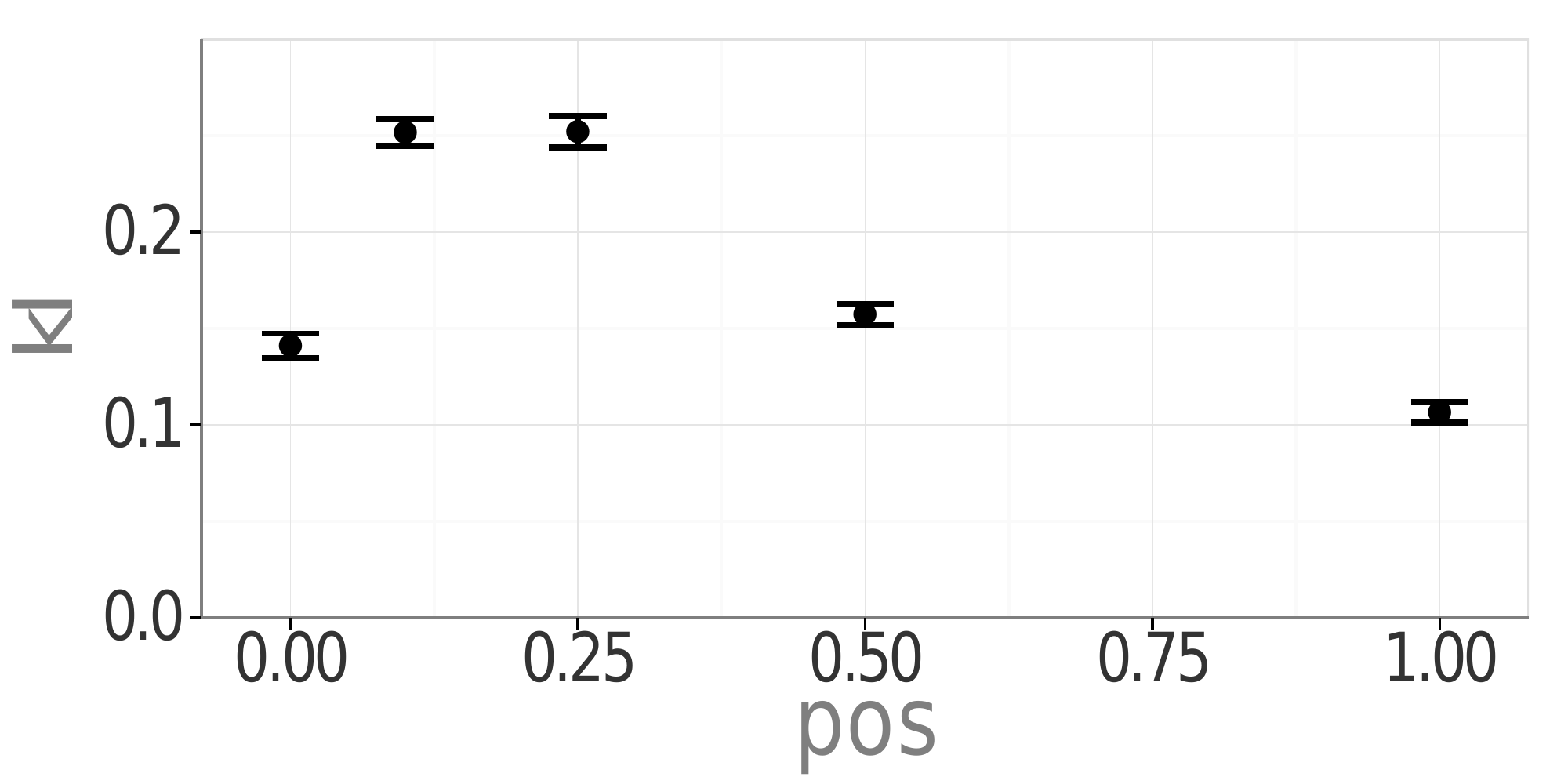}
    \vspace*{-.15in}
    \caption{$kl$ vs $pos$}
\label{f:trend_kl_maxpos}
  \end{subfigure}
  \begin{subfigure}[b]{0.24\textwidth}
      \includegraphics[width=1\columnwidth]{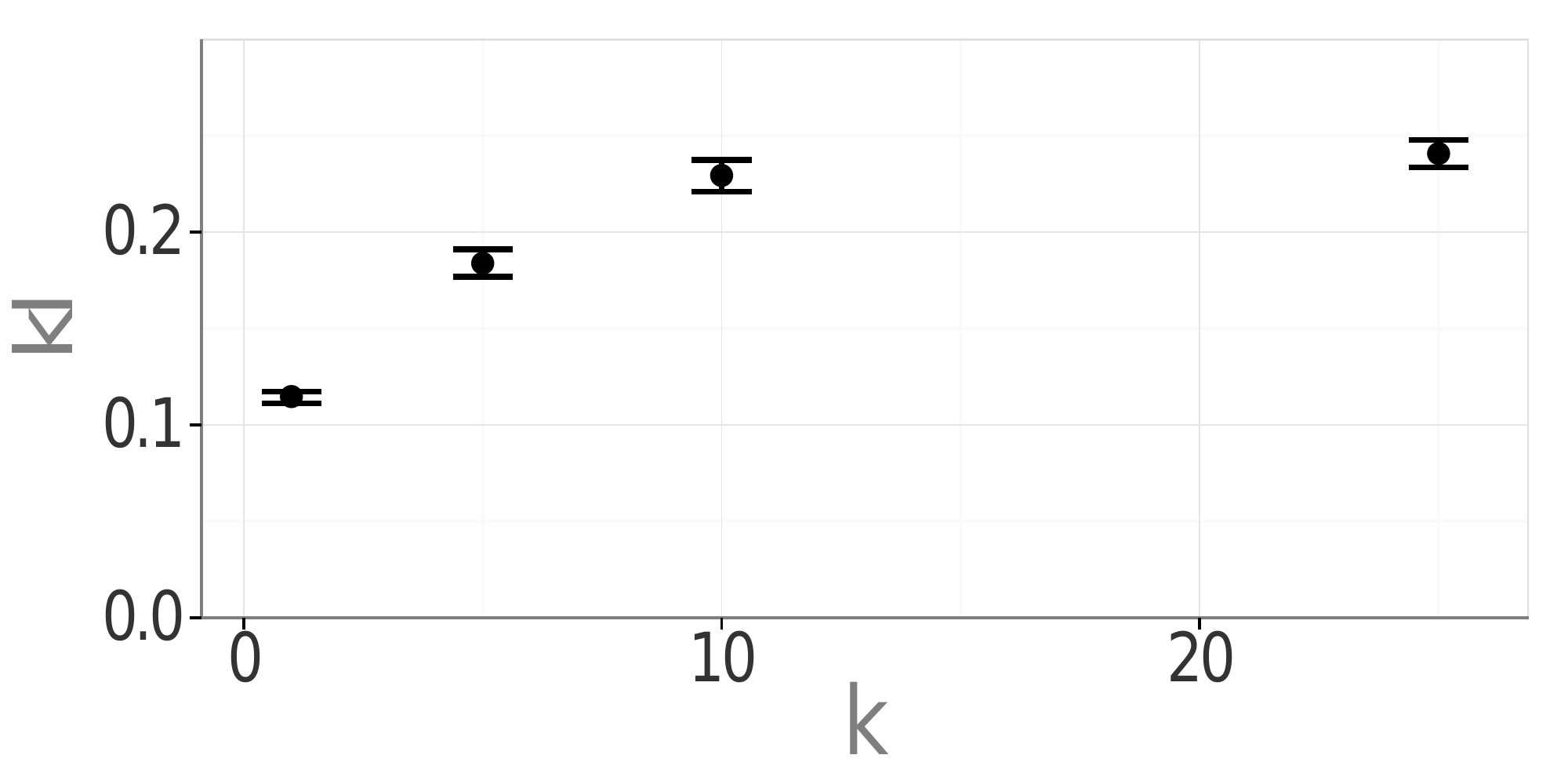}
    \vspace*{-.15in}
    \caption{$kl$ vs $k$}
  \end{subfigure}
  \begin{subfigure}[b]{0.24\textwidth}
      \includegraphics[width=1\columnwidth]{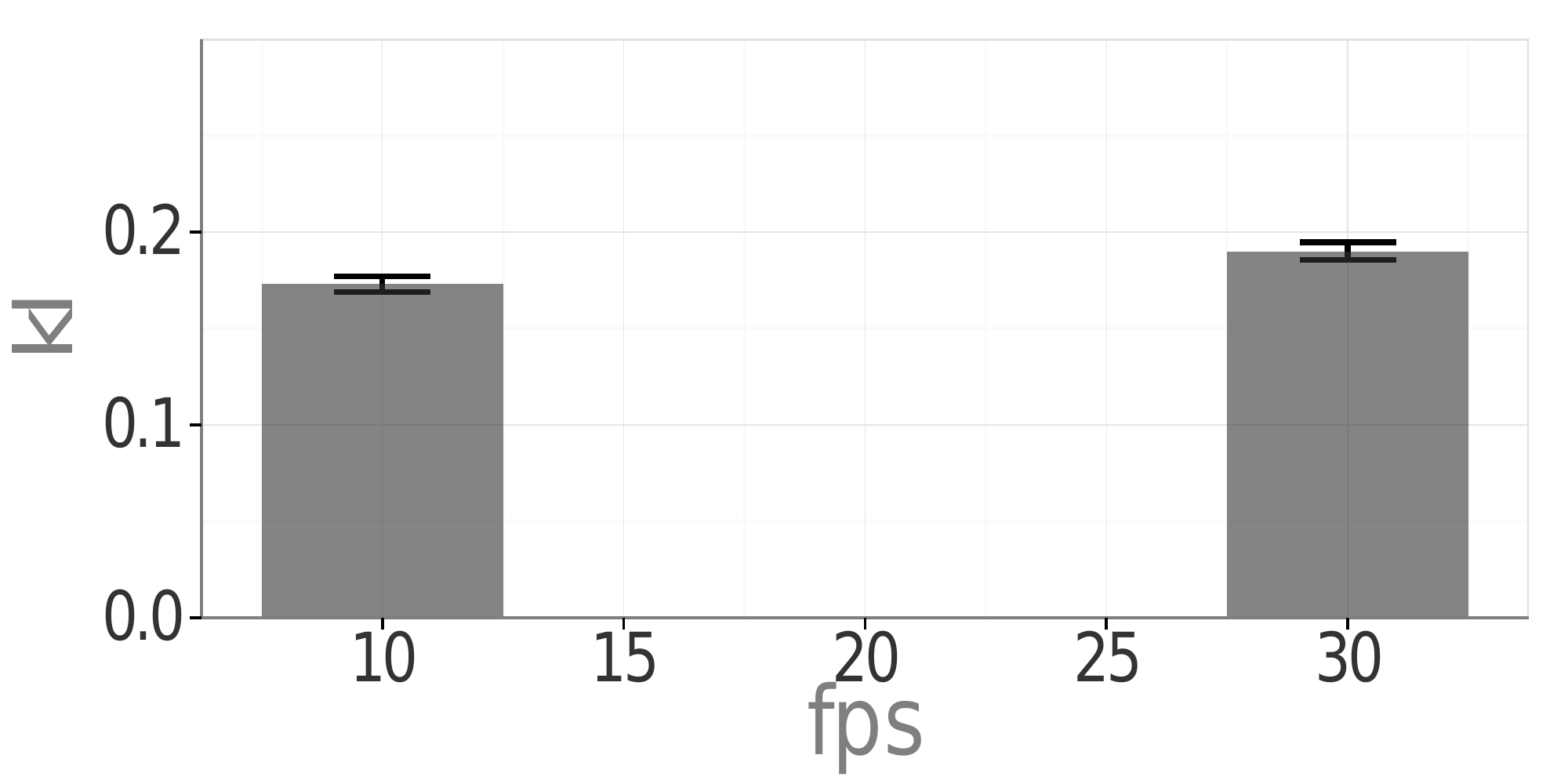}
    \vspace*{-.15in}
    \caption{$kl$ vs $fps$}
  \end{subfigure}
	\vspace*{-.1in}
  \caption{\emph{kl} distance between estimated and true trend under each factor}
  \label{f:trend_detect_kl}
\end{figure*}

\begin{figure*}[th!]
  \centering
  \begin{subfigure}[b]{0.3\textwidth}
    \includegraphics[width=\columnwidth]{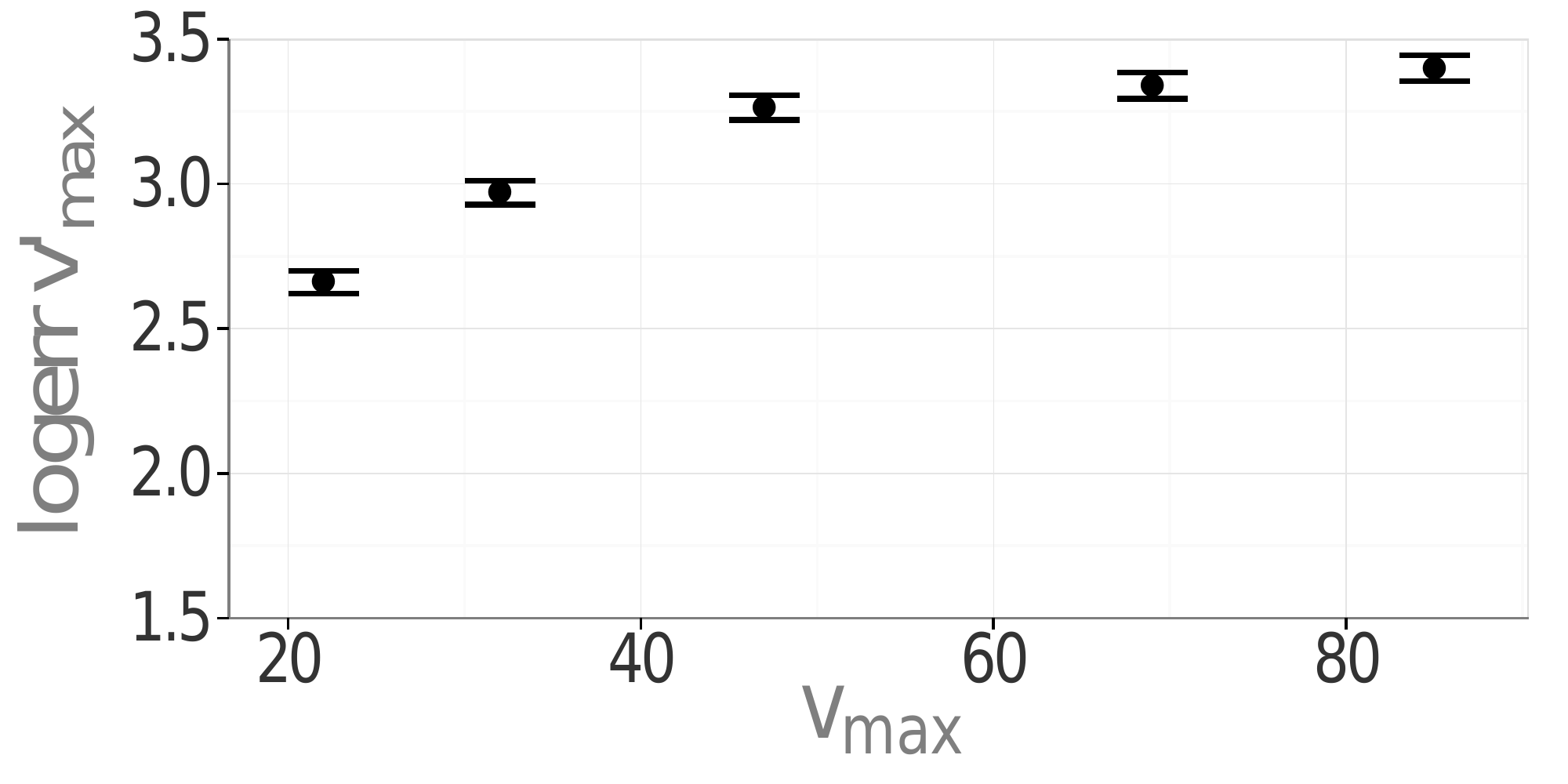}
    \vspace*{-.15in}
    \caption{Estimated max value ($v_{max}'$)}
  \end{subfigure}
  \begin{subfigure}[b]{0.3\textwidth}
    \includegraphics[width=\columnwidth]{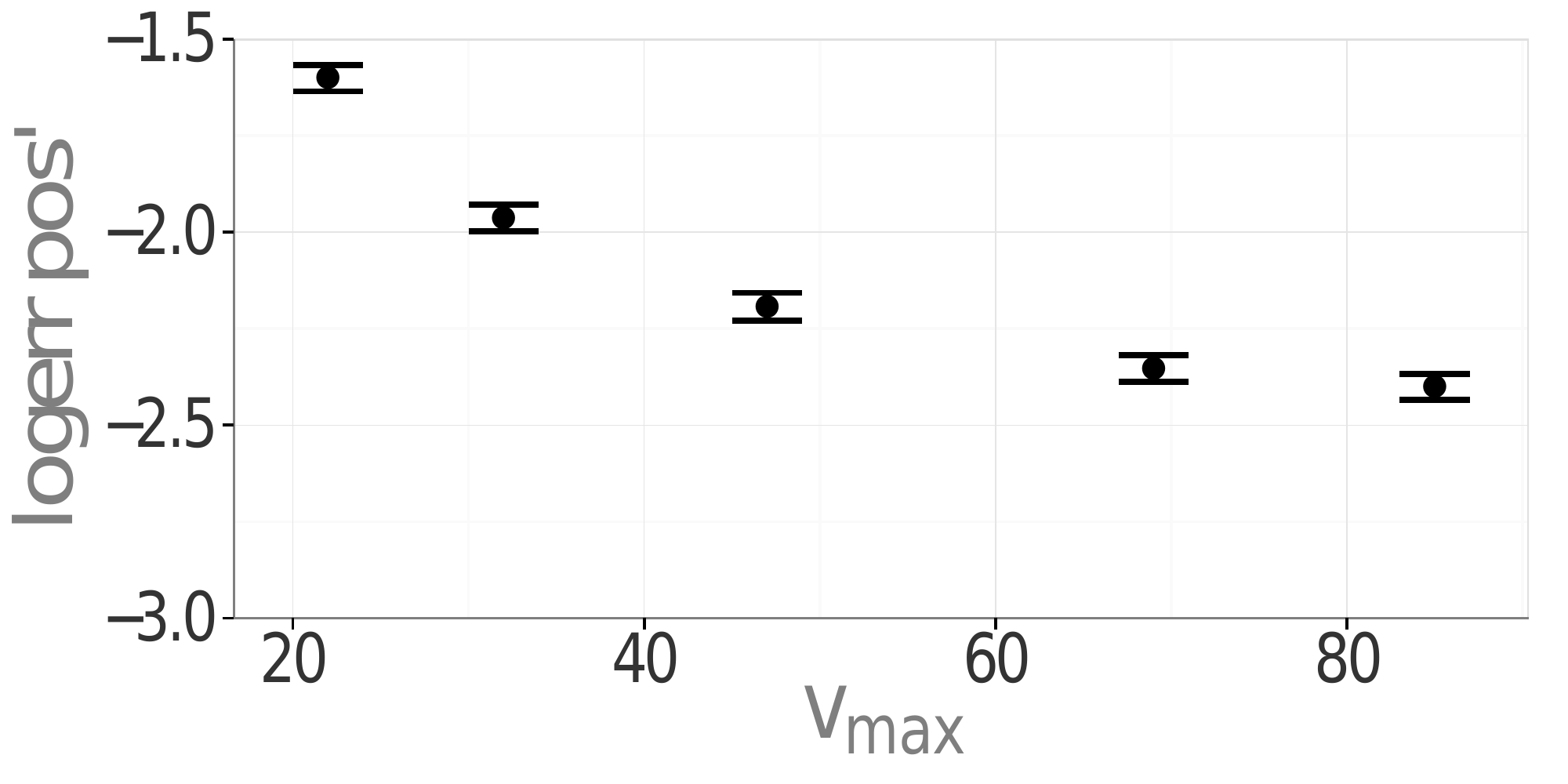}
    \vspace*{-.15in}
    \caption{Estimated position ($pos'$)}
  \end{subfigure}
  \begin{subfigure}[b]{0.3\textwidth}
    \includegraphics[width=\columnwidth]{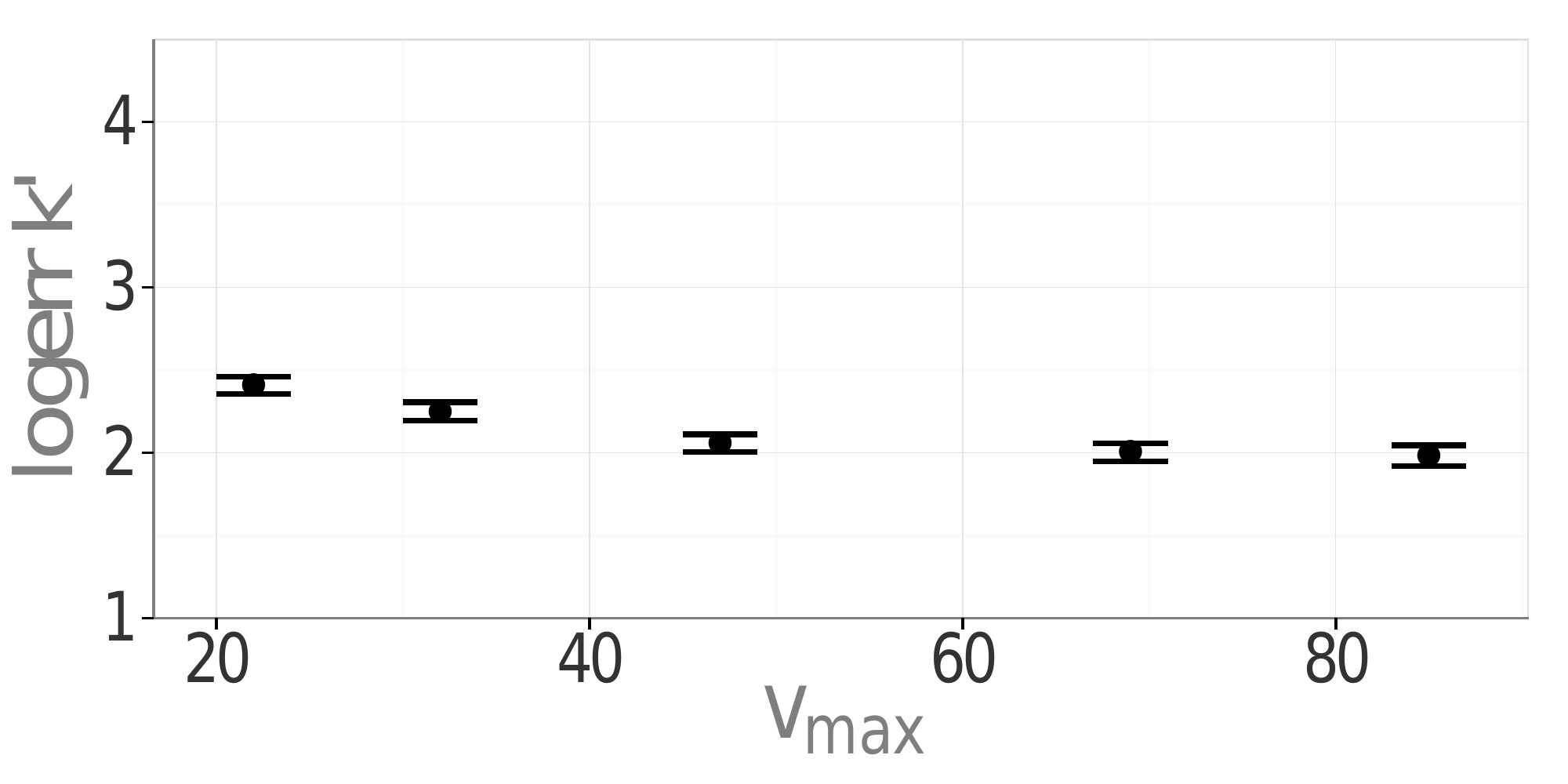}
    \vspace*{-.15in}
    \caption{Estimated rate of change ($k'$)}
  \end{subfigure}
	\vspace*{-.1in}
  \caption{Log error of user estimates for each factor as the true maximum value ($v_{max}$) varies.}
  \label{f:trend_max}
\end{figure*}

\begin{figure*}[th!]
\centering
  \begin{subfigure}[b]{0.3\textwidth}
    \includegraphics[width=\columnwidth]{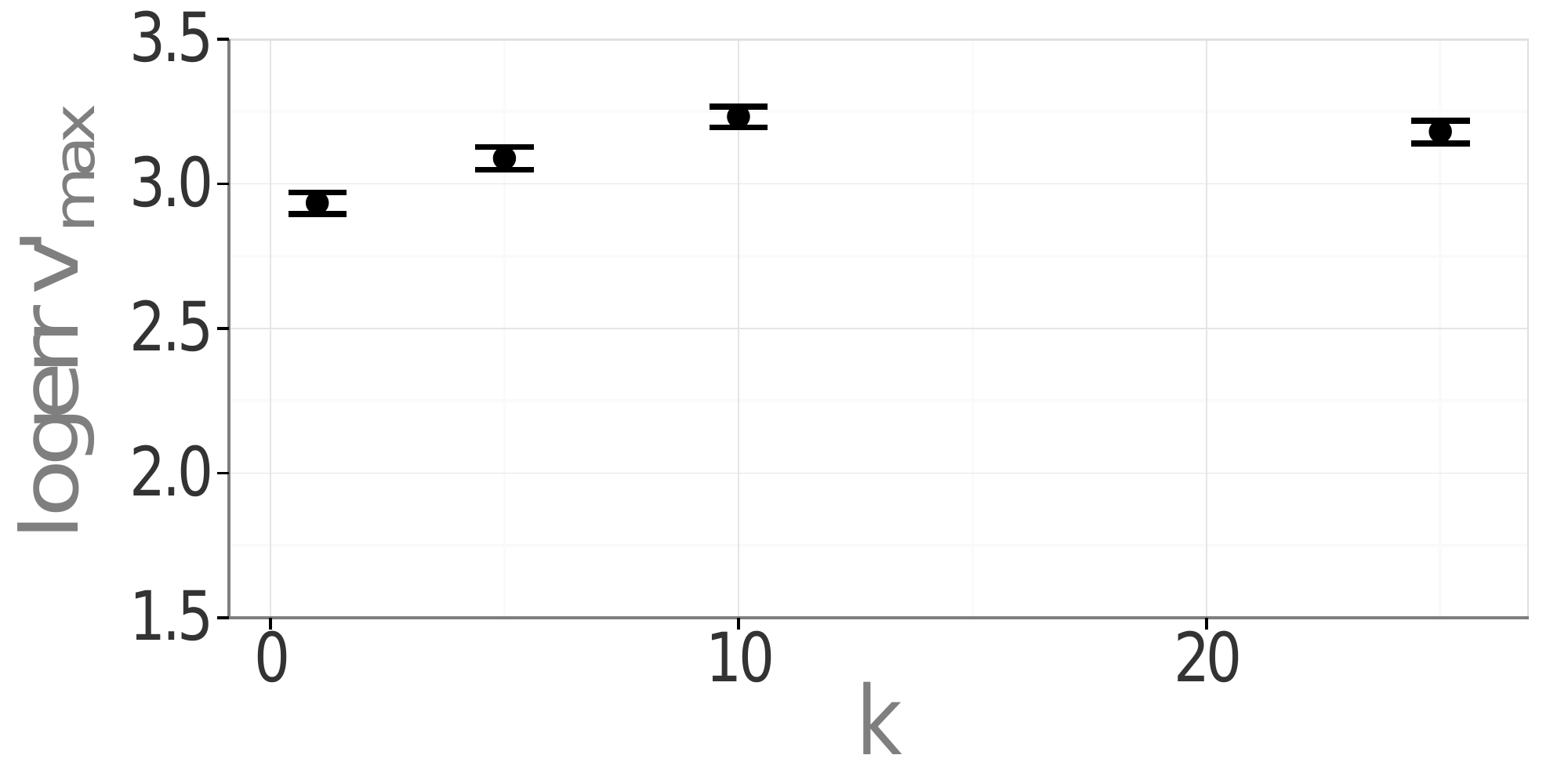}
    \vspace*{-.15in}
    \caption{Estimated max value ($v_{max}'$) vs $k$}
    \label{f:trend_max_k}
  \end{subfigure}
  \begin{subfigure}[b]{0.3\textwidth}
      \includegraphics[width=\columnwidth]{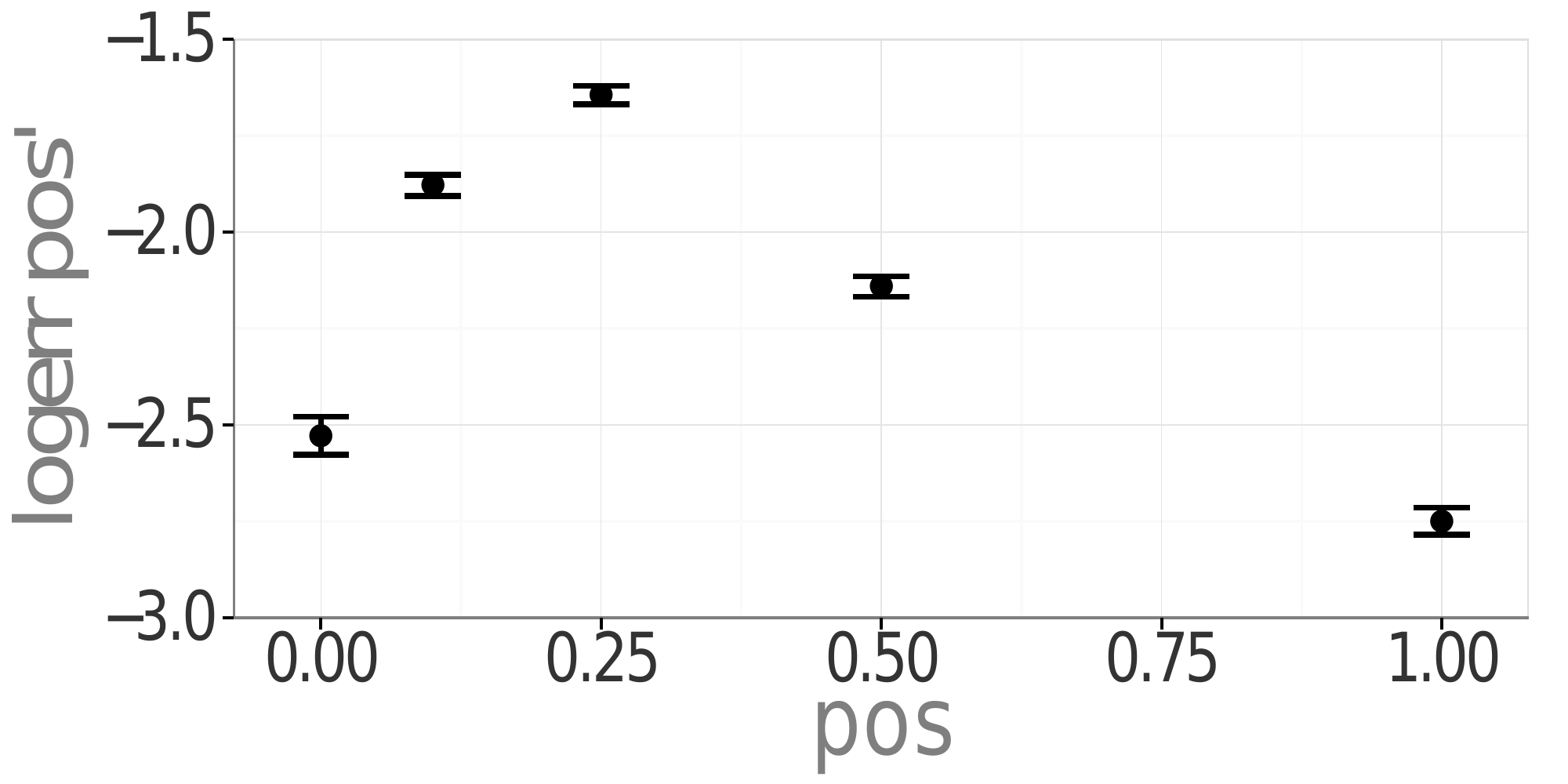}
    \vspace*{-.15in}
    \caption{Estimated position $pos'$ vs true $pos$ }
    \label{f:trend_pos_maxpos}
  \end{subfigure}
  \begin{subfigure}[b]{0.3\textwidth}
    \includegraphics[width=\columnwidth]{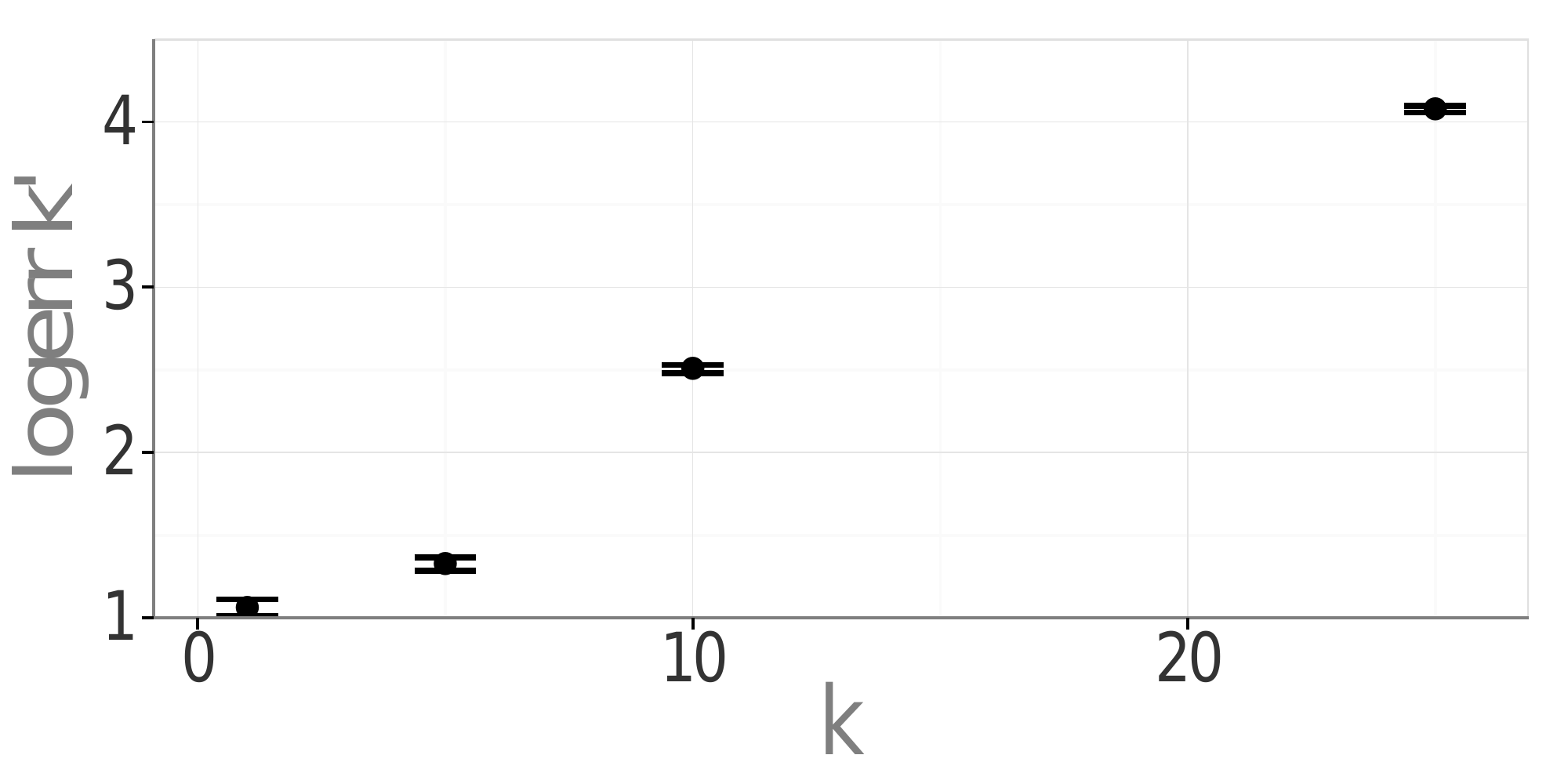}
    \vspace*{-.15in}
    \caption{Estimated rate of change ($k'$) vs $k$}
    \label{f:trend_k_k}
  \end{subfigure}
	\vspace*{-.1in}
  \caption{Log errors of user estimates for each factor as true position ($pos$) and the rate of change ($k$) varies.}
  \label{f:trend_posk}
\end{figure*}

We now turn to the complex trend estimation task (e.g., sales rapidly increased in the 3rd quarter).  
This task requires multiple simultaneous comparisons of the trend functions parameters $v_{max}$, $pos$ and $k$.
Our key goals are to understand:
\begin{CompactEnumerate}
\item What parameters most affect accuracy in this task, and how they compare with the sensitivity in the extrema task?
\item How does the estimate of the maximum value in the trend judgment compare with that in the extrema task?
\end{CompactEnumerate}

\subsection{Materials and Procedure}

We use the same materials and procedure as the extrema experiments.
The more complex task design require changes to the task interface so that users can input their estimates of the target bar's trend.
One option is to allow the user to directly draw a curve as part of the estimate, however this confounds the 
input with the user's expertise and control over the mouse.
Instead, we designed a slider-based interface (Fig.\ref{f:trendinterface}) to specify $k'$, $pos'$, and $v_{max}'$ estimates by interacting with sliders.  
The interface keeps a visualization of the user's estimated trend $\mathcal{T}'$ up to date (Figure~\ref{f:trendcontrols}).
A consequence of this interface is that the user must estimate three different attributes of the animation, including
the maximum value, the maximum position and rate of change.  One benefit of this design is that it affords us the opportunity
to compare the effect of task difficulty on the accuracy of $v_{max}$ estimates.

We use the Kullback-Leibler (KL) divergence~\cite{kullback1959statistics}, 
which varies from $0$ when the inputs are identical and increases as the inputs diverge, 
to estimate trend's similarity to the true trend $\mathcal{T}$.
We approximate KL by quantizing and comparing $\mathcal{T}$ and $\mathcal{T}'$ 
at each of the $380$ pixels:  
$kl(\mathcal{T}, \mathcal{T}') = \sum_{x=0}^{380} \mathcal{T}(\frac{x}{380}) \times ln(\frac{\mathcal{T}(\frac{x}{380})}{\mathcal{T}'(\frac{x}{380})})$

The input interface is more complex, so we used three qualification tasks and three training tasks.  
The qualification tasks ensure users are able to use the input controls---a static curve is overlaid on
the input interface and users must replicate the static curve to withn $kl<0.01$ before they can proceed.  
The training tasks are identical to a real task---users estimate the trend of an animation.
On submission, it overlays the true trend over the user's estimate and encourages the user to re-adjust their submission until satisfaction. 

The extrema experiments showed that both data and animation parameters affect user judgement accuracy,
while color marking is definitively more accurate.
As such, we marked the target bar using color and designed a broad parameter sweep
(\texttt{Trend} in Table~\ref{t:params-wide}) resulting in a $200$ factor design.
A large number of judgements is likely to overwhelm users~\cite{lorist2000mental}, so we randomly 
selected $100$ judgments for a given participant, and ran $190$ assignments.

\subsection{Results}

We remove spammers and outliers by separately applying the filtering procedure from Section~\ref{s:extrema-results} 
to each of the user's submissions of $k'$, $pos'$, $v_{max}'$.  We remove the result if it is rejected by any of the filters.
We found that the existing correlation threshold ($0.8$) was so stringent that it removed all submitted judgements.
This is partly because the task is more difficult---the average judgment time $10$s is nearly twice the time for extrema judgements,
only $90$ of $132$ assignments were fully completed, and the correlation was low for all participants.
In response, we reduced the correlation threshold to $0$, and ultimately analyzed $58$ filtered, completed assignments.
Our findings are robust to the specific thresholds.

We use a linear model to analyze the $kl$ metric, and the same log-linear model as the extrema experiments for the other user estimates.

\smallskip
\noindent\textbf{Overall Trend Results:} 
Figure~\ref{f:trend_detect_kl} compares the KL distance $kl$ and each of the input parameters.
Similar with the extrema results, $kl$ increases as the true maximum $v_{max}$, rate of change $k$, or frame rate $fps$ increase.
Interestingly, $kl$ is minimized when the target maximum is in the first or last frame (e.g., strictly linear trend functions),
and is maximized when the position is early in the animation $pos\in[0.1,0.25]$.
This relationship is in contrast to the extrema experiments (Figure~\ref{f:logerr_maxpos}) where 
judgement was \emph{least accurate} when $pos=0$;  we study this in depth below.

KL-divergence is a useful summary statistic, however it implicitly assigns importance to the 
accuracy of each of the estimated data parameters.
To better understand the sources of the $kl$ plots, we analyzed
how judgements of each parameter are affected by their true values.

\noindent\textbf{Dependence on $v_{max}$: }
Figure~\ref{f:trend_max} shows how the user estimates depend on the true maximum value.
Although the log error of $v_{max}'$ increases with the true maximum, as in the extrema experiments, 
the curve is shifted up---the absolute error is $\sim1.65\times$ larger under the same experimental conditions.
We hypothesize this shift is due to the user's shift of focused attention~\cite{healey2007perception,healey2012attention} from specific value of the target bar to its distribution.
It is possible that the $v_{max}'$ estimates in the trend experiments may degrade because users estimate using preattentive rather than cognitive processing 
due to the user's additional focus on the positional and rate of change aspects of the animation.
In contrast, increasing $v_{max}$ reduces log errors of maximum position and rate of change estimates by nearly $2$ and $0.4$, respectively.
It is possible that the larger value improves the ease of tracking the target bar.

\begin{figure}[h]
\centering
  \includegraphics[width=\columnwidth]{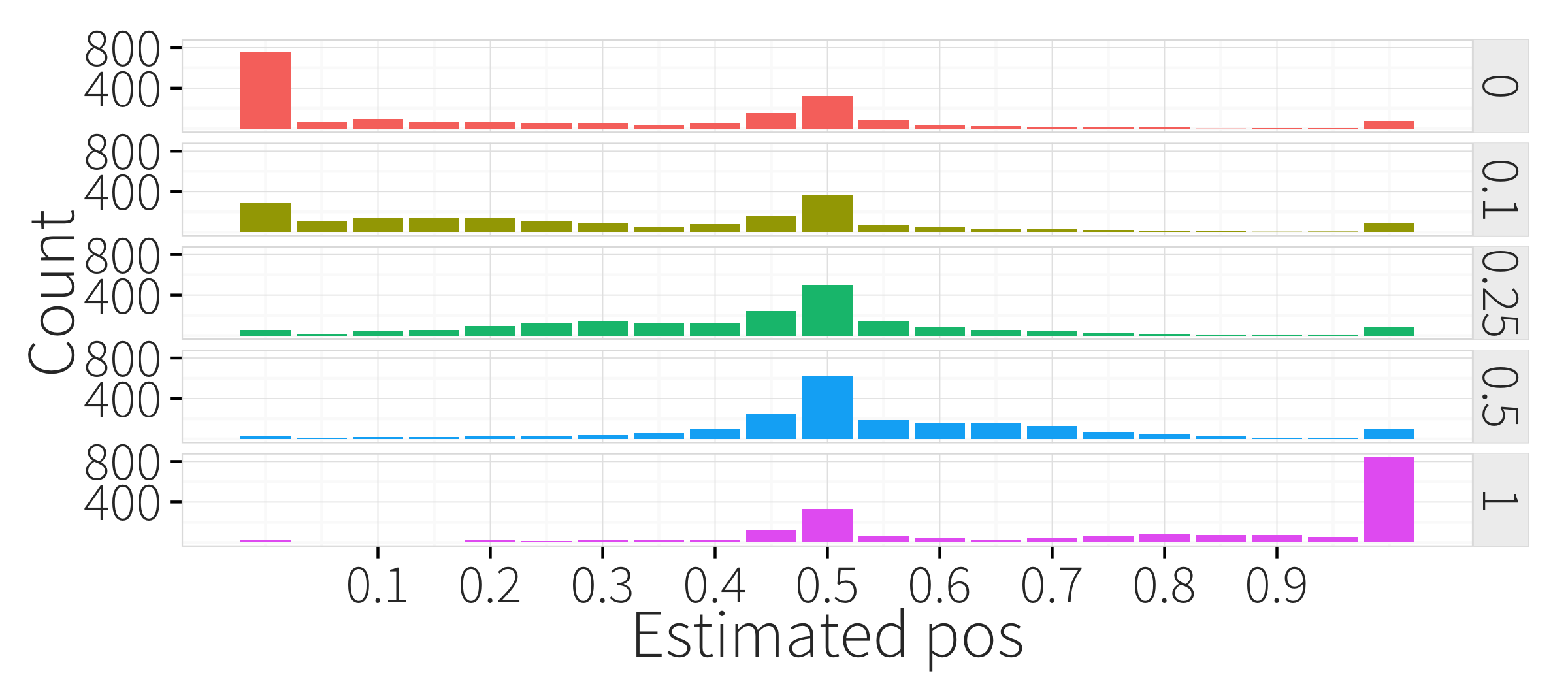}
  \vspace*{-.15in}
  \caption{Histogram of $pos'$ estimates in buckets of $0.05$, conditioned on true $pos$.}
  \label{f:trend_pos_histogram}
  \vspace*{-.1in}
\end{figure}

\noindent\textbf{Dependence on Other Parameters:}
We found that the log errors were not dependent on $pos$ and $k$ except in three cases, depicted in Figure~\ref{f:trend_posk}.
The first is the dependence on the true maximum position $pos$.  We find that neither $v_{max}$ nor $k'$ depended on the maximum position;
in contrast, the log error of the estimated $pos'$ in Figure~\ref{f:trend_pos_maxpos} is nearly identical to the pattern in Figure~\ref{f:trend_kl_maxpos}, suggesting a causal relationship.
To better understand the shape of this curve, we plotted the histogram of $pos'$ estimates for each of the true $pos$ values (Figure~\ref{f:trend_pos_histogram}).  
Rather than rounding effects~\cite{talbot2014}, which are less likely due to the slider interface, we found considerable bias towards $0$, $0.5$ and $1$.  
For example, when the true $pos \in \{0.1, 0.25, 0.5\}$, the number of estimates within $0.05$ of 
the three values constituted $44.6\%$ of all judgments.  
Some users appear to estimate the true position based on the slight increase in estimates around the true position ($pos=0.1, 0.25$),
however this is dwarfed by the prevalence of the biased estimates.

The second case is the slight relationship between $v_{max}'$ and the true rate of change $k$ shown in Figure~\ref{f:trend_max_k} ($\chi^2(1,N=5739)=15.7, p=7e^{-5}, \beta_{k}=0.009$).
It appears to converge at a similar point as the color curve in the extrema experiments (Figure~\ref{f:logerr_exponent}), and lower than the circle curve.
It is not clear if this is exhibiting a baseline accuracy in this judgement task, or if it is a side effect of the experimental design.
In the last case, it is very difficult for users to accurately estimate how quickly the target bar 
changes $k'$---the absolute error increase by up to $8\times$ as compared to estimating $k'$ for linear rates of change.

\noindent\textbf{Discussion:}
Our analysis used KL divergence as a proxy for overall accuracy, and finds that distribution characterization
of a target bar strongly depends on all four data and animation parameters.
Interestingly, the relationship between $kl$ and the maximum position is not monotonically increasing, and exhibits a ``bump'' when $pos\in(0, 0.5]$.
This shape may be explained by the relationship between the log error of $pos$ and the true value of the maximum position (Figure~\ref{f:trend_pos_maxpos}).

For the most part, we find that increasing a given data parameter (e.g., $pos$) predominantly affects user judgement of 
the same parameter, and has minor effect on the other estimated parameters.  
For example, changes in the maximum position has little effect on the accuracy of $v_{max}$ and $k$.
In contrast, increasing $v_{max}$ \emph{improves} user judgements of $pos$ and $k$, possibly because changes in the target bar are magnified and thus easier to perceive. 

On average, shifting the user's task to distribution charecterization reduces the 
absolute error of the extrema judgement by nearly $1.65\times$ as compared to the previous experiment that
expilcitly judges the target bar's extrema value.

These results show that user perception is highly dependent of the specific data values being animated,
as well as the specific aspects of the trend---the overall trend or one of its parameters.
One possible use of this result is that, if we know which aspect of the trend the user is focusing on, 
there may be utility in exaggerating other aspects of the animation to improve the specific task's perceptual accuracy.
For example, in an exploratory setting, a user may first look for candidate patterns by 
scanning for very rapid changes, and once found, then focus on quantifying values in those patterns.  
This type of visualization manipulation is similar to fish-eye~\cite{sarkar1992graphical} techniques
in interactive visualization, however more studies are needed to understand the utility and
whether there are unforseen side-effects.

\section{Limitations and Future Work}
\label{sec:discussion}

To the best of our knowledge, this is the first study that quantitatively measures graphical perceptual accuracy in animated data visualizations.  As such, there are a number of limitations that warrant further investigation in order to bridge the gap between our current study and perceptual evaluations in the context of live data visualization interactions. Some of these limitations are due to our choices in experimental design, while we believe others are fundamental to animation and interaction oriented perceptual studies.  

One limitation is simply due to our attempt at deriving quantitative results from a finite sample size---given the vast variety of human perceptual systems, backgrounds, experience with data visualizations, viewing conditions, fatigue and other contextual difference, it is difficult to make blanket statements about human perception in general.  These are all candidates to study in isolation, however we believe it is still helpful to identify general patterns that may help inform design.  Going further, we are excited about the potential for personalized perceptual models that can help tailor interactive interfaces to an individual's characteristics (e.g., color sensitivity or eyesight).


A second limitation was our choice to fix the animation length while varying the frame rate.  By doing so, we were able to control for memory effects. An alternative would be to fix the number of frames and allow the animation length to vary with the frame rate. Both approaches rely on picking a constant (animation length vs \# of frames) and it is unclear how this choice would affect our results.

Third, we focused on two simple judgment tasks---extrema and trend distribution---in the context of animation.  However, in practice, users may be performing any number or combination of judgement tasks such as comparisons of pairs of marks, or judging multiple visual encodings at the same time (e.g., color and position).  We might hypothesize that the user is characterizing the distribution when she rapidly scrubs a scroll bar, however further studies are needed to better predict users' high level goals~\cite{horvitz98lumire} -- particularly when the goals may be unclear to the user~\cite{hutchins1985direct,Carver02controlprocesses}.  Understanding the user's intended tasks will continue to be a valuable direction of work.

Finally, there is still a large gap between the animation-oriented experiments in this paper and understanding the graphical perception of interactions.  We made simple assumptions to decompose interactions into short animation units, and evaluated individual animations.  However, there is still considerable  work to generalize these findings to longer, more complex animations, and finally to interactions.

\section{Conclusion}
\label{sec:conclusion} 

This paper extends graphical perception studies from judgements of static data visualizations
to judgments of animated visualizations.  We proposed a simple model that decomposes simple
direct manipulation interactions (e.g., brushing) into a sequence of 
elementary animated perceptual chunks that can be parameterized and used in reproducible experiments.
We then studied variations of these chunks for two simple judgment tasks---reading
the maximum value and the distribution of a target bar throughout the animation.

Our observations verify some known results and also present several new insights.
Consistent with the tracking literature, denoting the target using color rather than shapes
 drastically improves the user's ability to make quantitative judgements about the target bar.
Although larger values improve the ability to detect the rate and timing of changes in an animation,
they also increase the error of the estimated bar height.
In contrast to values, estimating \emph{when} the target is maximal is extremely hard to judge.
In addition, timing plays an important role: changes earlier in the animation were found to be harder to perceive as 
compared to changes in the middle or end of an animation.

We believe our work is a promising step towards principled research that combines perception and interactive data visualization systems. 
Our findings impact two complementary aspects of animated and interactive visualizations spanning both the HCI and systems communities. 
First, just as the results of static perceptual experiments have been used towards automatic visualization recommendation systems 
such as APT~\cite{apt}, ShowMe~\cite{mackinlay2007show}, and Voyager~\cite{voyager}, 
we expect our insights to influence how encodings and animation parameters are selected for animated visualizations of a dataset. 
Second, we believe there is tremendous potential for concrete uses beyond measurements of efficacy between various visual encodings.
For example, we hope to embed extensible perceptual models as part of the data visualization system in order to enhance
performance while remaining aware of perceptual limitations.
In concert, by understanding the user's sensitivity to various data and animation parameters, we
can better design interactive visualizations that are usable, accurate, and performant.

\clearpage
\newpage
\bibliographystyle{SIGCHI-Reference-Format}
\bibliography{main}
\end{document}